\documentclass[twoside]{article}

\usepackage[accepted]{aistats2020}

\usepackage[utf8]{inputenc} 
\usepackage[T1]{fontenc}    

\usepackage{booktabs}       

\usepackage{hyperref}       
\usepackage{url}            

\usepackage{multicol} 

\usepackage{afterpage}
\usepackage{graphicx}
\graphicspath{{img/}} 

\usepackage{algorithm}
\usepackage{algorithmic}

\usepackage{amsmath,amssymb,amsthm,amsfonts} 

\theoremstyle{definition}

\theoremstyle{remark}

\DeclareMathOperator*{\argmin}{arg\,min}

\usepackage{tikz}
\usetikzlibrary{cd} 
\usetikzlibrary{backgrounds}

\usetikzlibrary{fit,positioning,arrows.meta,bayesnet,shapes.geometric,chains,matrix,scopes,calc,decorations.markings}

\tikzstyle{param}=[circle, minimum size = 0.7cm, thick, draw=black!100, fill = gray!10, node distance = 0.5cm]
\tikzstyle{data}=[rectangle, minimum size = 0.7cm, thick, draw =black!100, node distance = 0.5cm]
\tikzstyle{model}=[rectangle, minimum size = 1cm, thick, draw=black!100, node distance = 0.5cm]

\usepackage{natbib} 
\usepackage{breakcites} 

\usepackage{Sweave}
\begin{document}

\runningtitle{Semi-Modular Inference}

\runningauthor{Carmona and Nicholls}

\twocolumn[

\aistatstitle{Semi-Modular Inference: enhanced learning in multi-modular models by tempering the influence of components}

\aistatsauthor{
Chris U. Carmona \\
\And
Geoff K. Nicholls \\
}

\aistatsaddress{ Department of Statistics\\
University of Oxford\\
Oxford, UK \\
\texttt{carmona@stats.ox.ac.uk}\\
\And
Department of Statistics\\
University of Oxford\\
Oxford, UK \\
\texttt{nicholls@stats.ox.ac.uk}\\
}
]

\begin{abstract}
Bayesian statistical inference loses predictive optimality when generative models are misspecified.

Working within an existing coherent loss-based generalisation of Bayesian inference, we show existing Modular/Cut-model inference is coherent, and write down a new family of \emph{Semi-Modular Inference (SMI)} schemes, indexed by an influence parameter, with Bayesian inference and Cut-models as special cases. We give a meta-learning criterion and estimation procedure to choose the inference scheme. This returns Bayesian inference when there is no misspecification.

The framework applies naturally to Multi-modular models. Cut-model inference allows directed information flow from well-specified modules to misspecified modules, but not vice versa. An existing alternative power posterior method gives tunable but undirected control of information flow, improving prediction in some settings. In contrast, SMI allows \emph{tunable and directed} information flow between modules.

We illustrate our methods on two standard test cases from the literature and a motivating archaeological data set.
\end{abstract}

\section{Introduction} \label{sec:intro}

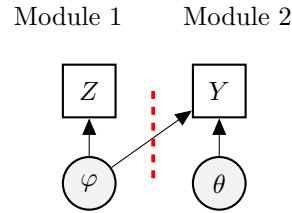
\begin{figure}[!ht]
  \begin{center}
  \begin{tikzpicture}
      \node (Z) [data] {$Z$};
      \node (Y) [data, right=of Z, xshift=0.5cm] {$Y$};
      \node (phi) [param, below=of Z] {$\varphi$};
      \node (theta) [param, below=of Y] {$\theta$};
      \edge {phi} {Z,Y};
      \edge {theta} {Y};
      \draw[dashed,red,line width=0.5mm] (0.85,0) to (0.85,-1.25);
      \node[text width=3cm] at (0.5,1) {Module 1};
      \node[text width=3cm] at (2.75,1) {Module 2};
  \end{tikzpicture}
  \end{center}
  \caption{Graphical representation of a simple multi-modular model.}
  \label{fig:toy_multimodular_model}
\end{figure}

Consider statistical inference in a multi-modular setting. The model for the available data is assembled from several component \emph{modules}. Each module describes a probabilistic relation between observable variables (data) and unknown quantities (parameters, latent variables, missing data). Modules may share parameters, missing data and other latent variables. Fig.~\ref{fig:toy_multimodular_model} illustrates a simple two-module model with a shared parameter $\varphi$ (ignore the dashed line explained in Sec.~\ref{sec:cut_models}). The first module has data $Z$ and one parameter, $\varphi$, while the second module has data $Y$ and two parameters, $\theta$ and $\varphi$.

In conventional Bayesian inference, parameters are jointly informed by the data and model assumptions shared across the modules. As a consequence, large-scale multi-modular analyses are particularly susceptible to problems arising from model misspecification, as any bad module may distort inference in the model as a whole \citep{Liu2009}.

This drawback has motivated alternative inferential approaches that modify conventional Bayesian learning in order to regulate feedback between modules. \emph{Modular} inference \citep{Liu2009, Jacob2017b} also known as a \emph{Cut-model} inference \citep{Spiegelhalter2014, Plummer2015} completely eliminates the contribution from some modules to the posterior distribution of parameters in other modules (see Section~\ref{sec:cut_models}).
However, we may do better to moderate, rather than eliminate, the influence of misspecified modules.

We give a new \emph{Semi-Modular Inference (\textbf{SMI})} which \emph{smoothly} regulates the influence of modules on the overall inference. The procedure effectively expands the space of \emph{candidate distributions} \footnote{following \cite{Jacob2017b}, we refer to any distribution representing beliefs on parameters $\theta$ (or $\varphi$, or both) as a \emph{candidate distribution} for the parameters} in such a way that Bayesian inference and Cut-model inference are particular cases.

When it comes to expanding the inference framework, an ``anything goes'' approach is clearly suspect. We stay within the class of inference schemes defined by \cite{Bissiri2016}. Those authors define and characterise coherent loss-based inference and note that Bayesian inference and the power posterior \citep{Walker2001} are coherent. We set out existing Cut-model and Power-posterior inference in Sections~\ref{sec:cut_models} and~\ref{sec:power_lk}, introduce SMI in Section~\ref{sec:smi} and then describe the encompassing framework of \cite{Bissiri2016} in Section~\ref{sec:gen_inf}, where we show that SMI and Cut-model inference are coherent.

The SMI posterior interpolates candidate distributions between the conventional Bayesian posterior and the Cut-model posterior. Candidate distributions are indexed by a continuous \emph{degree of influence} parameter $\eta$. This controls the contribution of a module to the candidate distribution. When $\eta$=0 the candidate distribution is the Cut-model posterior and when $\eta$=1 it is the Bayesian posterior. The SMI posterior is not a scheme for model elaboration with an extra parameter. SMI-based inference with any value of $\eta$ other than $\eta=1$ is not Bayesian inference.

In Section~\ref{sec:data_analyses}, we apply SMI to model-based inference for simulated and real-world datasets and evaluate its performance. It is easy to understand why it outperforms Bayesian and Modular inference in these examples. The supplementary material provides proofs and additional numerical experiments. All results are reproducible using the accompanying R package\footnote{\url{https://github.com/christianu7/aistats2020smi}}.

\section{Background methods}\label{sec:background}

\subsection{Modular Inference: cut model}
\label{sec:cut_models}

The Cut model is an alternative to Bayesian inference designed to remove unwanted feedback from poorly specified modules. The OpenBUGS manual \citep{Spiegelhalter2014} describes the cut function with the words ``The cut function acts as a kind of valve in the graph: prior information is allowed to flow downwards through the cut, but likelihood information is prevented from flowing upwards''. Cut model inference is a form of Bayesian multiple imputation.

Consider again the two-module configuration of Fig.~\ref{fig:toy_multimodular_model}. In standard joint or ``full'' Bayesian inference, information from the two modules informs every parameter, so in particular the values of $Y$ will in general influence the posterior distribution of $\varphi$. The posterior distribution of $(\varphi,\theta)$ in Bayesian inference is
\begin{equation} \label{eq:full_01}
p(\varphi,\theta \mid Z,Y) = p(\varphi \mid Z,Y) p( \theta \mid Y, \varphi ) .
\end{equation}
The marginal distribution of $\varphi$ depends on $Y$.

Now, suppose that for some reason (usually because we suspect some model misspecification) we want the parameter $\varphi$ to learn only from module 1, \emph{cutting} the influence from module 2 on $\varphi$. The cut is represented in Fig.~\ref{fig:toy_multimodular_model} by a dashed line on the edge from $\varphi$ to $Y$; it denotes an inference structure in which $\varphi$ influences $Y$ but not vice versa (following \cite{Lunn2009}).

Under this modified scheme, the Cut-model ``posterior'' for  $(\varphi,\theta)$ is
\begin{equation} \label{eq:cut_01}
p_{cut}(\varphi,\theta \mid Z,Y) = p(\varphi \mid Z) p(\theta \mid Y,\varphi).
\end{equation}
Notice that the marginal distribution of $\varphi$ no longer depends on $Y$.

Cut-model inference is a form of Bayesian Multiple Imputation \citep{Lunn2009, Styring2017} in which we make multiple imputation of $\varphi$ and then analysis of $\theta$ given the imputated distribution of $\varphi$. The literature identifies potential advantages: cut models may simplify inference \citep{Cox1975}; prevent unwanted feedback from suspect models \citep{Lunn2009}; improve MCMC mixing \citep{Plummer2015}; reduce the MSE in estimates \citep{Liu2009}; increase predictive performance \citep{Jacob2017b}; answer the need in some settings to make a sequential analysis in which the data $Z$ is not shared with the analyst carrying out inference for $\theta$.

\subsection{Power posterior} \label{sec:power_lk}

In the power posterior we raise the likelihood to a power, seeking to improve robustness under model misspecification (\cite{Walker2001, Bissiri2016, Holmes2017, Grunwald2017, Miller2018a}).

Consider independent data $Y=(Y_1,\ldots,Y_n)$ generated from an unknown true distribution $f^*(Y)$. Assume we have a data model $f(Y|\theta)$ and a prior distribution $p(\theta)$. For a fixed $\eta \in \mathbb{R}$, we define the \emph{$\eta$-powered posterior} $p_{pow, \eta}(\theta|x)$ as
\begin{equation} \label{eq:power_lik_01}
p_{pow, \eta}(\theta|Y) = \frac{ f(Y|\theta)^\eta p(\theta) }{p_{\eta}(Y)} \\
\end{equation}
with $p_{\eta}(Y) = \int f(Y|\theta)^\eta p(\theta) d\theta$ the \emph{powered} normalising constant.

The new parameter $\eta$ is called the \emph{learning rate}, following \cite{Grunwald2012}. The learning rate calibrates the influence of the prior relative to that of the data; if $\eta \in [0,1]$ the prior is given more influence and the data less. When $\eta>1$ the data is given more prominence, and in the extreme case when $\eta$ is very large the posterior accumulates around the maximum likelihood estimate for the model. For the misspecification we encounter, we tend to be interested in the case $\eta \in [0,1]$.

A key point emphasised by \cite{Grunwald2017} among others is that this is not simply model elaboration. We do not put a prior on $\eta$ and learn it in the usual Bayesian way. Roughly speaking the learning rate ``corrects'' Bayesian inference and should not be chosen using Bayesian inference but according to other ``external'' criteria, for example, a predictive loss on test data. \cite{Grunwald2012} and \cite{Grunwald2017} propose the \emph{SafeBayes} algorithm to find the optimal learning rate. In that work, the learning rate $\eta$ is chosen to maximise the “sequentially randomised” Bayesian marginal log-likelihood. This can be interpreted as measure of predictive accuracy. In contrast \cite{Holmes2017} choose the learning rate by matching the prior expected gain in information between the prior and posterior. This gain in information is quantified by the expected divergence in Fisher information.

\section{Semi-Modular Inference} \label{sec:smi}

In this section, we define \textbf{\emph{Semi-Modular Inference (SMI)}}, a modification of Bayesian inference in multi-modular settings which allows us to adjust the flow of information between data and parameters in separate modules.
Referring to the two-module example in Fig.~\ref{fig:toy_multimodular_model}, we allow the data from module 1 to dominate in inference for $\varphi$ without entirely discarding the joint structure provided by the full model.

Our approach is motivated by the observation in \cite{Plummer2015} that cut-model inference is Bayesian multiple imputation: we might expect to do better at the second analysis stage of cut-model inference if we can more accurately impute missing values in the first stage. A two stage analysis resembling Cut-model analysis, but using a power posterior in the first stage delivers this. First, we update our beliefs about $\varphi$ using a power likelihood, with power $\eta \in [0,1]$ on module 2. The power-posterior improves Bayesian multiple imputation of $\varphi$ at the expense of $\theta$. In the second stage, we re-learn our beliefs on $\theta$ conditional on the learnt distribution of $\varphi$.
 The \emph{degree of influence}, $\eta$, controls the contribution of the suspect module in the inference.

\subsection{SMI distributions} \label{sec:smi_def}

Let $p(Z | \varphi)$ and $p(Y | \varphi,\theta)$ denote the observation models for the two modules. We introduce an auxiliary parameter $\tilde\theta$, expanding the model parameters from $(\varphi,\theta)$ to $(\varphi,\theta,\tilde\theta)$.

We define the \textbf{$\eta$-smi posterior} as
\begin{equation} \label{eq:smi_02}
 p_{smi,\eta}(\varphi,\theta,\tilde\theta|Z,Y) = p_{pow,\eta}(\varphi,\tilde\theta|Z,Y) p(\theta|Y,\varphi)
\end{equation}
where $p_{pow,\eta}( \varphi , \tilde\theta \mid Z, Y )$ is the power posterior
\begin{equation}
  p_{pow,\eta}( \varphi , \tilde\theta \mid Z, Y ) \propto p(Z|\varphi) p( Y \mid \varphi, \tilde \theta )^\eta \;  p(\varphi,\tilde\theta).
\end{equation}
Expanding in terms of model elements (details in supplement),
\begin{align*} \label{eq:smi_03}
 p_{smi,\eta}(\varphi,\theta,\tilde\theta|Z,Y) \propto & \; p(Z \mid \varphi) \; p(Y \mid \varphi, \tilde\theta )^{\eta} \; p(Y \mid \varphi, \theta ) \\ 
 & \times\quad \frac{1}{ p(Y \mid \varphi) } \; p(\varphi, \theta, \tilde\theta),
\end{align*}
where $p(Y \mid \varphi) = \frac{1}{p(\varphi)} \int p(Y \mid \varphi, \theta ) \; p(\varphi, \theta) d\theta $.

The $\eta$-smi posterior of the original parameters is just the marginal,
\begin{equation}\label{eq:smi_marg}
p_{smi,\eta}(\varphi,\theta|Z,Y)=\int p_{smi,\eta}(\varphi,\theta,\tilde\theta|Z,Y) d\tilde\theta.
\end{equation}

The posterior distribution $p_{smi,\eta}( \varphi,\theta | Z,Y )$ interpolates between the Bayesian posterior and the Cut model posterior.
When $\eta=1$ the SMI posterior is the usual Bayesian posterior (Eq.~\ref{eq:full_01}),
\begin{align*}
p_{smi,1}(\varphi,\theta | Z,Y ) &= p( \varphi |Z,Y )p( \theta |Y, \varphi) \\
&= p( \varphi,\theta |Z,Y ),
\end{align*}
whereas if $\eta=0$, the SMI posterior of $\varphi$ gives back the Cut model (Eq.~\ref{eq:cut_01}),
\begin{align*}
p_{smi,0}(\varphi,\theta | Z,Y ) &= p( \varphi | Z )p( \theta |Y, \varphi) \\
&= p_{cut}( \varphi,\theta |Z,Y ).
\end{align*}

Semi-modular inference is defined for a fixed degree of influence $\eta \in [0,1]$. Each value of $\eta$ yields a different \emph{candidate distribution}, $p_{smi,\eta}$, which we call the SMI posterior, representing posterior belief on $(\varphi,\theta)$. Natural questions now are, how to choose in a principled manner the ``best'' degree of influence, and how and why does SMI help? The answer to the latter question is in a sense straightforward. If a generalised inference scheme achieves a better score, according to some agreed external criterion, we should use it, and not otherwise. This approach is taken in \cite{Jacob2017b}. We answer the first question in the next section.

\section{Analysis with (Semi-)Modular Inference} \label{sec:smi_considerations}

In this section we show that inference with the SMI posterior distribution at fixed $\eta$ is valid (in the sense of \cite{Bissiri2016}) and give an MCMC algorithm targeting $p_{smi,\eta}(\varphi,\theta|Z,Y)$. We give criteria and estimation procedures for choosing
$\eta$, and comment on relations with cut-models and the power posterior.

\subsection{Coherence of (Semi-)Modular Inference} \label{sec:gen_inf}

We apply the general framework for updating belief distributions in \cite{Bissiri2016} to show that the SMI posterior - and hence also the cut posterior - are \emph{valid} and \emph{coherent} updates of beliefs.
This framework is based on a loss function $l(\theta;y)$ connecting information in the data to the parameters of interest. The log-likelihood is one such loss, but \cite{Bissiri2016} give examples where other losses may be relevant, and Cut-models and SMI prove to be further examples.

\cite{Bissiri2016} characterise a valid belief update. They list a number of axiomatic requirements. We verify that our SMI-update satisfies these axioms in the supplement. The most demanding of these conditions, in our setting, is the coherence condition.

In \emph{Coherent} inference we reach the same posterior distribution, whether we update belief using all data simultaneously or update belief taking the data sequentially in independent blocks. In our two-module setting, this applies in several ways: we can observe responses from different modules one after the other (e.g. first $Z$, and then $Y$); we can observe sequential data fragments within the same module (e.g. first $Z_1$, and then $Z_2$, with $Z=(Z_1,Z_2)$); any mixture of these.

The generalised update of belief in \cite{Bissiri2016} follows a decision theoretic approach. In single module notation, the generalised posterior distribution $p_{l_\rho}$ arising from a loss $l_\rho(\theta;y)$ in a family of loss functions indexed by $\rho$, is the probability measure minimising a cumulative loss function
$L_{\rho}(\nu;p_0,Y)$ over choices of probability measure $\nu$,
\begin{equation*}
p_{l_\rho}(\theta \mid Y) = \argmin_{\nu} L_{\rho}(\nu;p_0,Y).
\end{equation*}
The cumulative loss function, $L_{\rho}(\nu;p,Y)$ balances the expected loss in the fit to data and the Kullback-Leibler divergence from the posterior to the prior distribution (generically $p_0$ say), and is defined by
\begin{equation*}
L_{\rho}(\nu;p_0,Y)=\int l_{\rho}(\theta,Y) \nu(d\theta) + d_{KL}(\nu,p_0).
\end{equation*}
\cite{Bissiri2016} show that the optimal, valid and coherent update of beliefs from prior to posterior is given by
\begin{equation*}
p_{l_\rho}(\theta \mid Y ) \propto \exp\{-l_\rho( \theta ; Y ) \} p_0(\theta)
\end{equation*}

The canonical case in the single-module setting is the logarithmic loss function $l(\theta;Y) = - \log f(Y \mid \theta) $, which yields the conventional Bayesian update of beliefs given by the posterior distribution. The power posterior is obtained by taking the loss function $l_{pow,\rho}(\theta;Y) = - \rho \log f(Y \mid \theta)$.

In the Supplementary material we prove that, for the model in Fig.~\ref{fig:toy_multimodular_model}, the loss function which yields the Cut model posterior is
\begin{align} \label{eq:cut_loss}
  l_{cut}( (\varphi,\theta) ; (Z,Y) ) =& - \log p(Z \mid \varphi) \\
  &-\log p(Y \mid \varphi,\theta) + \log p(Y \mid \varphi), \nonumber
\end{align}
and the loss function yielding the SMI posterior is
\begin{align} \label{eq:smi_loss}
  l_{smi,\eta}( (\varphi,\theta,\tilde\theta) ;& (Z,Y) ) = - \log p(Z \mid \varphi) \\ & -\eta \log p(Y \mid \varphi,\tilde\theta) \nonumber \\
  &-\log p(Y \mid \varphi,\theta) + \log p(Y \mid \varphi). \nonumber
\end{align}
The $p( Y \mid \varphi )$ terms in each expression are the loss-function expression of the idea of cutting feedback from $Y$ to $\varphi$.

We prove that both Cut-model inference and SMI are coherent when we update using the correct associated loss function given above. Detailed proofs are given in the supplementary material.

\subsection{Targeting the modular posterior} \label{sec:cut_mcmc}

\cite{Plummer2015} and \cite{Jacob2017b} explain that an MCMC algorithm that correctly targets the cut distribution cannot usually be implemented, due to the presence of the intractable normalising constant $p( Y \mid \varphi )$. A SMI sampler faces the same issue.

In order to sample the SMI-posterior in a single MCMC run we need, refering to Eq.~\ref{eq:smi_02}, a standard MCMC sampler for $p_{pow,\eta}(\varphi, \tilde\theta \mid Z, Y)$ and an \emph{exact} sampler for $p( \theta \mid Y,\varphi )$.
It is straightforward to check that the transition kernel given by a two-stage update using these two conditional distributions satisfies detailed balance. A proof is given in the supplement.

Exact simulation of $p( \theta \mid Y,\varphi )$ may be impracticable. In practice, there are currently three options, nested MCMC, unbiased MCMC via couplings \citep{Jacob2017}, and tempered transitions \citep{Plummer2015}.

\textit{Unbiased MCMC via couplings} simulates samples unbiased in expectation from the Cut-model posterior. The approach uses coupled pairs of Markov Chains sharing a common transition kernel, which almost-surely meet at some finite time $\tau \geq 1$ and stay together thereafter.
The same approach applies to SMI, though we have not implemented this.

In our examples we use \textit{nested MCMC}, described in Algorithm~\ref{alg:smi_nested_mcmc}: sample $N_1$ draws from $p_{pow,\eta}(\varphi, \tilde\theta \mid Z, Y)$; for each sampled value of $\varphi$, run a sub-chain targeting $p(\theta \mid \varphi,Z)$ for $N_2$ steps, where $N_2$ is large enough to avoid initialisation bias; keep only the last sampled value in this sub-chain. The resulting joint samples $(\varphi,\tilde\theta,\theta)$ are approximately distributed according to the SMI posterior. We typically ignore the output for $\tilde\theta$ as we target the marginal in Eq.~\ref{eq:smi_marg}. The validity of this algorithm relies on a double asymptotic regime in $N_1$ and $N_2$ \citep{Jacob2017b}.
This works well, with standard MCMC convergence checks, if convergence of the MCMC targeting $p(\theta \mid Y, \varphi)$ is rapid.

\begin{algorithm}[tb]
\caption{Nested MCMC for SMI posterior Eq.~\ref{eq:smi_02}} \label{alg:smi_nested_mcmc}
\begin{algorithmic}
\STATE \textbf{Input:} influence $\eta\in[0,1]$, Data $(Z,Y)=\{ (Z_i,Y_i) \}_{i=1}^{n}$; observational models $f(Z \mid  \varphi )$ and $f(Y \mid \varphi,\theta)$; prior $p(\varphi,\theta)$; run-lengths $N_1$ and $N_2$.\\[0.1in]
\STATE \textbf{Output:} Samples $\{(\theta^{(s)}, \varphi^{(s)}) \}_{s=1}^{N_1}$ distributed approximately according to the SMI posterior $p_{smi,\eta}(\varphi,\theta \mid Z,Y )$ with influence parameter $\eta$.\\[0.1in]
   \FOR{$s = 1,\ldots,N_1$}
		\STATE Sample $(\tilde\theta^{(s)}, \varphi^{(s)}) \sim p_{\eta-pow}(\varphi,\tilde\theta|Z,Y)$, using any standard sampler.
   \ENDFOR
   \STATE Let $\{\varphi^{(s)}\}_{s=1}^{N_1}$ be samples after burn-in and thinning
   \FOR{$s=1,\ldots,N_1$}
      \FOR{$r=1,\ldots,N_2$}
         \STATE Sample $(\theta^{(s,r)}) \sim p( \theta \mid Y, \varphi^{(s)} )$, using any standard sampler.
      \ENDFOR
   \STATE Let $\theta^{(s)}=\theta^{(s,N_2)}$ (final state)
\ENDFOR
   \RETURN $\{(\theta^{(s)}, \varphi^{(s)}) \}_{s=1}^{N_1}$
\end{algorithmic}
\end{algorithm}


\subsection{Choosing the influence parameter} \label{sec:opt_eta}

We work in a $\mathcal{M}$-open setting \citep{Bernardo2000},
as model misspecification is our motivation for Semi-Modular Inference.
Conventional Bayesian inference is optimal for prediction (for the objective defined below) under an idealised scenario of correct model specification, full availability of data, and no computational restrictions. In the $\mathcal{M}$-open setting the conventional posterior may be outperformed by another candidate distribution \citep{Jacob2017b}.

The class of SMI candidate posteriors is indexed by $\eta$, so we need to give a procedure to choose a belief update operation from the set of candidate models $\mathcal{M} = \{ p_{smi,\eta} ; \eta \in [0,1] \}$. Following \citep{Bernardo2000}, we should determine $\eta$ on the basis of expected utility, provided some utility function.

We consider \emph{out-of-sample predictive accuracy} of the model as our utility function. Our criterion is the \textit{expected log pointwise predictive density} or ``{\it elpd}\,'',
\begin{align} \label{eq:elpd}
  elpd(\eta) = \int\int &p^*(z,y) \cdot \nonumber \\
  &\log p_{smi,\eta}( z, y \mid Z,Y) dz dy,
\end{align}
where $p^*$ is the distribution representing the true data-generating process and
\begin{align*}
  p_{smi,\eta}(z,y \mid Z,Y)=\int\int & p(z,y \mid \varphi, \theta) \cdot \\
  & p_{smi,\eta}(\varphi,\theta \mid Y,Z)\, d\varphi\,d\theta
\end{align*}
is a candidate posterior predictive distribution, indexed by $\eta$. See \cite{Vehtari2016} for further discussion of this measure. We take the value, $\eta=\eta^*$ say, which maximizes the estimated $elpd$-function.

We expect this criterion to yield $\eta\simeq 1$ when there is no model misspeciication and the data are informative of parameters. The $elpd$ is a KL-divergence, up to a constant not depending on $\eta$. If the posterior distribution of the parameters concentrates on the true values, as it may when there is no model misspeciication, then $p_{smi,\eta}( z,y \mid Z,Y)$ coincides with $p^*(z,y)$ when $\eta\simeq 1$ and this choice will minimise the divergence, and maximise the $elpd$. This is supported by our experiments.

In practice, $p^*$ is unknown, so we use cross-validation and WAIC \citep{Watanabe2009} estimators to approximate the $elpd$ at a grid of $J$ values of $\eta$. We tried both $elpd$-estimators as a check, and found good agreement. Leave-one-out cross-validation is natural but expensive. Other estimators are available \citep{Vehtari2016} and we are exploring these.

\subsection{The computational cost of SMI}

We compare the computational cost, $W_{smi}$ say, of SMI inference (using nested MCMC at $J$ values of $\eta$ and determining $\eta^*$) to the cost, $W_{bm}$ say, of doing regular Bayes-MCMC on the original problem.

The first stage of Algorithm~1 uses the same MCMC updates as Bayesian MCMC and a similar target, so it costs about $W_{bm}=N_1$ units. The second stage uses the same $\theta$-update as Bayes-MCMC
so costs no more than $N_2W_{bm}$. This is not quadratic in $N_1$ as $N_2$ is chosen to ensure that the stage two sampler converges but then produces just one draw from its target. The $J$ nested MCMC runs determining $\eta^*$ parallelise essentially perfectly across cores, and estimation of the WAIC and $\eta^*$ is in our experience a fast output analysis, so the overall cost is about $W_{smi}=W_{bm}+N_2W_{bm}$.

A more careful analysis considering thinning of MCMC chains (run the second half of Algorithm~1 on thinned output from the first half) shows that an overall SMI-cost $W_{smi} =KW_{bm}$ with $K\simeq 10$ should typically by achievable. This is justified in more detail in the Supplementary Material where the mixing times of the chains are taken into account.

\subsection{SMI and the Power Likelihood}
SMI is a two-stage procedure, fitting a power likelihood for $\phi$ and $\tilde\theta$, and then recalibrating $\theta$ conditional on $\phi$. Does the second stage improve the inference or should we simply stick with $\tilde\theta$?
The answer depends on the purpose of the inference. If interest lies purely in estimation of $\phi$, we should stay with the power posterior, as the second stage has no effect on the posterior for $\varphi$. However, if we are interested in $\theta$ or in prediction of $Z$ or $Y$, the best SMI candidate posterior may have a bigger $elpd$ and be selected over the power posterior. In Sections~\ref{sec:biased_data} and~\ref{sec:agric_analysis} we give examples where SMI improves prediction of new data (both sections) and mean square error (Sec.~\ref{sec:biased_data}, on synthetic data).

\subsection{SMI and Bayesian Multiple Imputation}\label{sec:dilution}
In a Bayesian setting missing observations are unknown quantities inferred jointly with unknown parameters. However, in some circumstances, there is an advantage in adopting different models for imputation and analysis, a situation known as \textit{uncongeniality} \citep{Meng1994,Xie2016a,Little2002}. This leads to Bayesian Multiple imputation.
Since SMI reduces to the Cut-model at $\eta=0$, we improve on multiple imputation (according to our criterion) if our procedure gives $\eta>0$. Our analysis in Section~\ref{sec:agric_analysis} illustrates this.

SMI address a phenomenon known, in the sense of \cite{Knuiman1998}), as "dilution". This is associated with “imputation noise” from uncongenial models. This noise typically causes the analyst to \textit{shrink} the estimated effect of interest towards zero. Cut-model inference is multiple imputation and suffers from this problem. SMI tends to reduce imputation-noise and dilution. This is picked up in the examples below.

\section{Examples} \label{sec:data_analyses}

Here we present three reproducible examples. In the first two cases, candidate distributions made available by Semi-Modular Inference outperform conventional Bayesian inference and the Cut-model. In the last, the Cut-model, or Bayesian inference are selected. These are special cases of SMI, so the extended inference is doing its job and returning the inference schemes with the best predictive performance.


\subsection{Simulation study: Biased data} \label{sec:biased_data}

This is a simple synthetic example in which the source of the ``misspecification'' is a poorly chosen prior. Suppose we have two datasets informing an unknown parameter $\varphi$. The first is a ``reliable'' small sample $Z=(Z_1,\ldots,Z_n),\ Z_i\sim N( \varphi, \sigma_z^2 )$, iid for $i=1,...,n$ distribution, with $\sigma_z$ known; the second is a larger sample $Y=(Y_1,\ldots,Y_m), Y_i\sim N( \varphi + \theta , \sigma_y^2 )$ iid for $i=1,...,m$, with $\sigma_y$ known. The ``bias'' $\theta$ is unknown.

This model was used by \cite{Liu2009} and \cite{Jacob2017b} as an example where modular/Cut-model approaches improve on Bayesian inference. We show that Semi-Modular Inference outperforms these inference schemes (which are special cases).

We choose true parameter values in such a way that each dataset offers apparent advantages to estimate $\varphi$. One dataset is unbiased but has a small sample size, $n=$25, whereas the second has an unknown bias but more samples, $m=$50, and smaller variance. Suppose the true generative parameters are $\varphi^*=$0, $\theta^*=$1, and we know $\sigma_z=$2 and $\sigma_y=$1. We assign a constant prior for $\varphi$, while $\theta$ is subjectively assessed to have a $N(0, \sigma_{\theta}^2)$ prior. We are over-optimistic about the size of the bias and set $\sigma_\theta=$0.5.

\begin{figure}[!ht]
\begin{center}
   \includegraphics[width=0.48\textwidth]{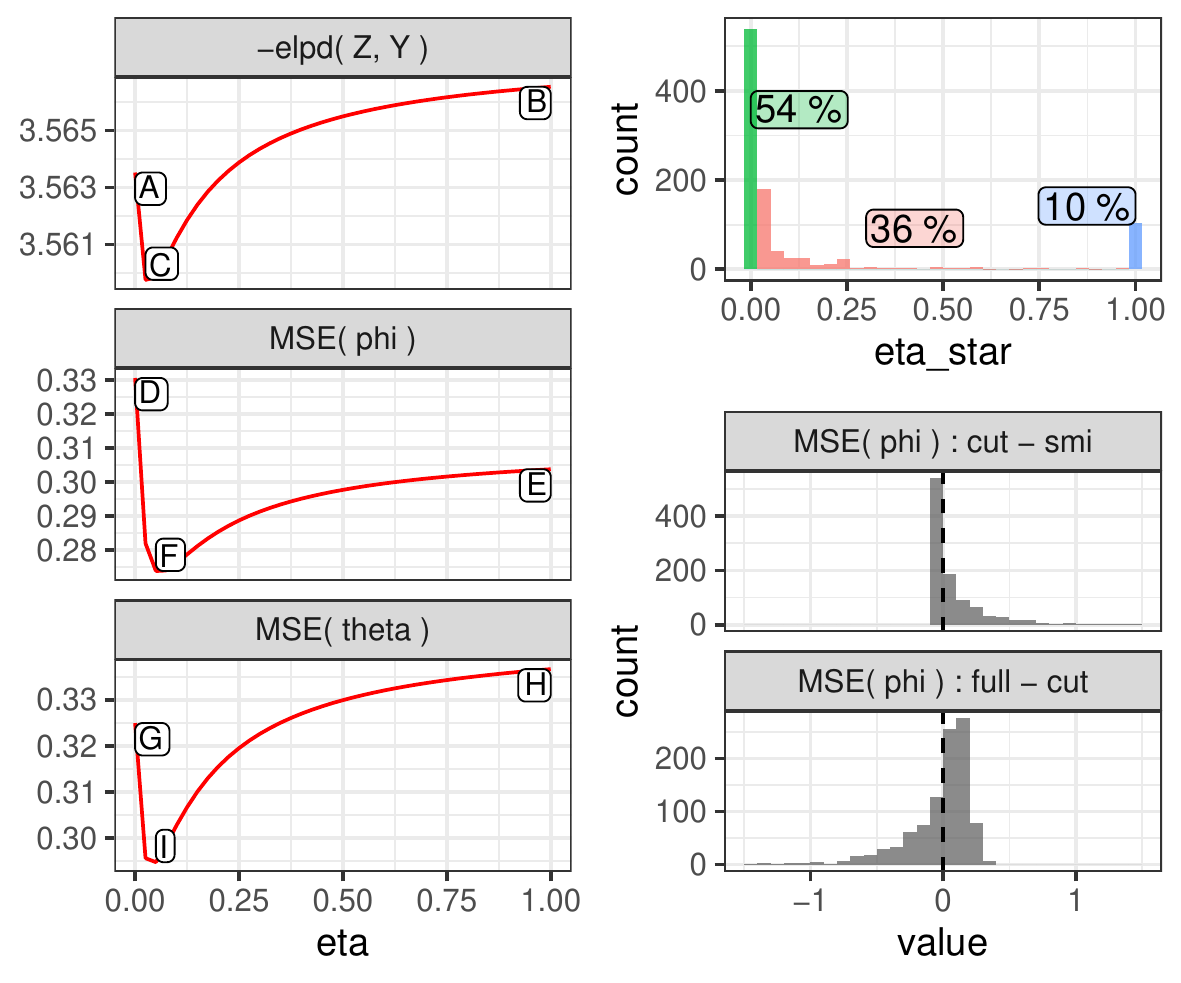}
\end{center}
   \caption[Biased data]{Model assessment for biased data example. Left column: average $-elpd$ and MSE for estimates under the SMI posterior with $\eta\in[0,1]$. Top-right: Histogram of optimal $\eta^*$ chosen for 1000 synthetic datasets. Bottom-Right: Histograms of differences in MSE between smi, cut and full-bayes.}
   \label{fig:SMI_biased_data}
\end{figure}

We calculate the SMI posterior and predictive distributions for each $\eta\in[0,1]$. The data are synthetic, so we calculate the Squared Errors (SE) $(\hat\varphi-\varphi^*)^2$ and $(\hat\theta-\theta^*)^2$ between the posterior mean and truth, and $-elpd$ as measures of performance. In the left column of Fig.~\ref{fig:SMI_biased_data} we display these metrics for $\eta\in[0,1]$, averaged across 1000 synthetic datasets.

As noted in \cite{Liu2009} and \cite{Jacob2017b}, the cut model posterior outperforms full-Bayes on average elpd (Fig.~\ref{fig:SMI_biased_data}: point A<B). SMI offers new candidate distributions which (on average) outperform the cut model and full-Bayes on prediction (point C v.s. A,B), estimating $\varphi^*$ (F v.s. D,E), and estimating $\theta$ (I v.s. G,H). The histogram of optimal $\eta^*$-values, for each synthetic dataset, is displayed top-right in Fig.~\ref{fig:SMI_biased_data}. In 36\% of the synthetic datasets, $\eta^*$ is not zero (cut model) or one (full-bayes). The two histograms at lower right show the distribution of the SE differences over datasets (the mean is the MSE difference), for estimating $\varphi^*$: SMI at $\eta^*$ against cut model (middle right), and cut model against full-Bayes (bottom right). SMI outperformed the cut model in 46\% of datasets (smaller SE), equal in 53.8\%, and beaten in 0.2\%, while the cut outperformed full-bayes in 61.7\% of datasets. See the supplement for further details.

\subsection{Agricultural data} \label{sec:agric_analysis}

In our second example, we apply SMI to the agricultural dataset introduced by \cite{Styring2017}, and analysed using a Cut-model. The authors test for specific agricultural practices in the first urban centres in Mesopotamia. Details of the model are given in \cite{Styring2017}. We give a brief outline here with more detail in the supplement, including a graphical representation.

The observation model has a normal response, $Y$, and regression parameters, variances and random effects we collect together as a parameter vector $\psi$. It has a three-level categorical variable ``manure-level'' $M$ with the same dimension as $Y$. Manure-level is missing data in roughly half the observations. The generative model for the missing values in $M$ is a proportional odds model with intercept parameters $(\alpha_1,\alpha_2)$ and the effect $\gamma$ for a higher-level covariate, ``site size''. Bayesian analysis suggests that the proportional odds module is misspecified. The parameter of scientific interest is $\gamma$; the authors test for $\gamma<0$. In our notation $M$ plays the role of $\varphi$ above, and $\gamma$ plays the role of $\theta$.

\begin{figure}[!ht]
\begin{center}
   \includegraphics[width=0.42\textwidth]{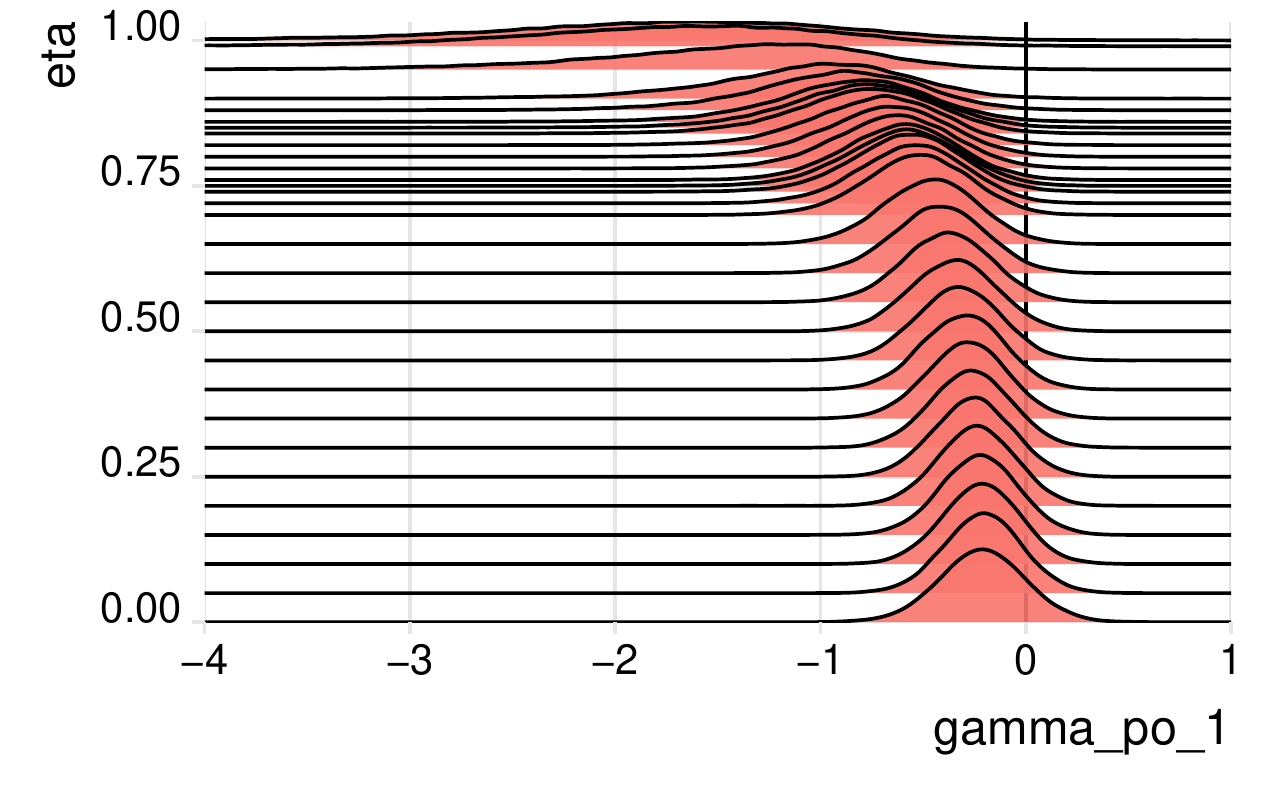}
\end{center}
   \caption{ Posterior distribution of $\gamma$ in the \emph{PO} module for different values of $\eta \in [0,1]$ }
   \label{fig:agric_smi_post_gamma}
\end{figure}

\begin{figure}[!ht]
\begin{center}
   \includegraphics[width=0.4\textwidth]{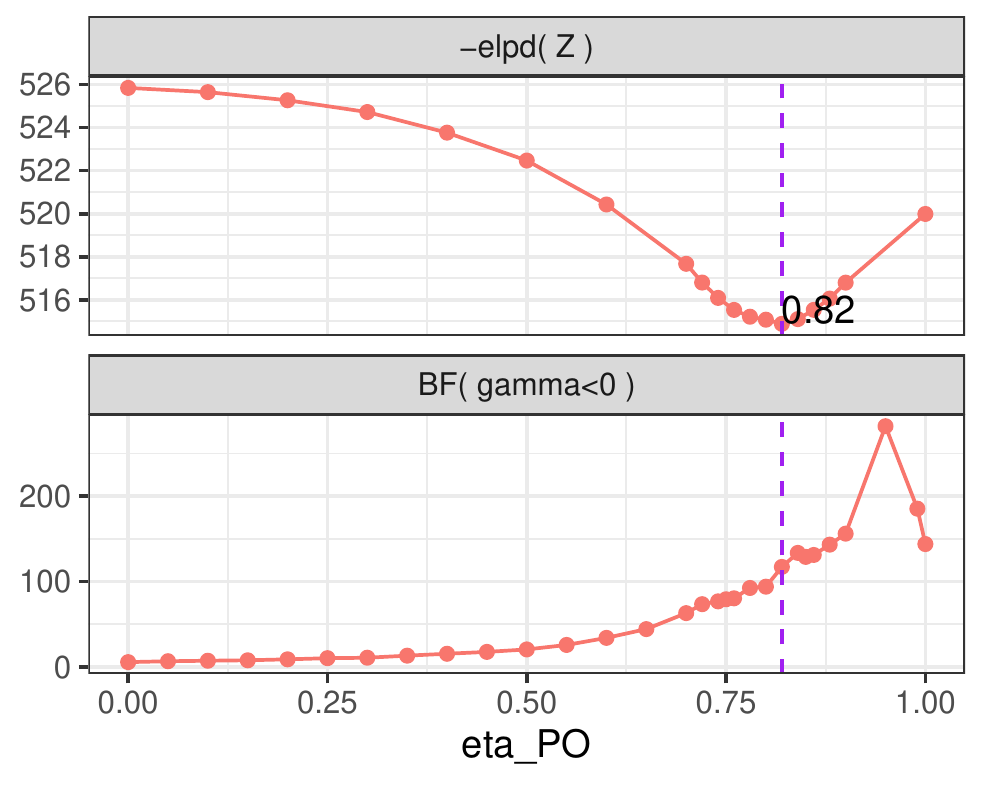}
\end{center}
   \caption{ Top: Estimated elpd as predictive criteria for choosing the value of $\eta \in [0,1]$ The maximum is reached near to $\eta=0.8$. Bottom: The Bayes Factor for the hypothesis $H_o:\gamma<0$ }
   \label{fig:agric_elpd}
\end{figure}

In Fig.~\ref{fig:agric_smi_post_gamma} we plot density estimates for the posterior distribution of $\gamma$ at a grid of values of $\eta$. The cut posterior is at the bottom ($\eta=0$). We can see the effect of dilution (see Section~\ref{sec:dilution}) as the mean $\gamma$ drifts towards zero as $\eta$ approaches zero. The Bayesian posterior is at the top ($\eta=1$). The estimated $elpd$ is plotted as a function of $\eta$ in the top panel of Fig.~\ref{fig:agric_elpd}. The $\eta$-value minimising the negative $elpd$ is 0.82. The evidence for $\gamma<0$ is much stronger at the candidate posterior as it suffers less \emph{dilution} than the cut posterior. This is quantified in the lower graph in Fig.~\ref{fig:agric_elpd} where we plot the posterior odds for $\gamma<0$ against $\eta$. These odds are the Bayes Factor (BF) at $\eta=1$, because the prior for $\gamma$ is symmetric about zero. We see the evidence for $\gamma<0$ is far stronger at the selected $\eta$-value (BF=117.11) than it is at the cut-model (BF=5.54).

\subsection{Epidemiological data} \label{sec:hpv_analysis}

In our final example, we apply SMI to an epidemiological dataset introduced by \cite{Maucort-Boulch2008}, studying the correlation between human papilloma virus (HPV) prevalence and cervical cancer incidence, revisited by \cite{Plummer2015} and \cite{Jacob2017b} in the context of cut vs full Bayes models.

The model has two modules: in each population $i=1,...,13$, a Poisson response for the number of cancer cases $Y_i$ in $T_i$ women-years of followup, and a Binomial model for the number $Z_i$ of women infected with HPV in a sample of size $N_i$ from the $i$'th population,
\begin{gather*}\label{eq:HPV_model}
Y_i \sim Poisson( \mu_i ) \\
\mu_i=T_i \exp( \theta_1+\theta_2 \phi_i ) \nonumber \\
Z_i \sim Binomial(N_i, \phi_i ). \nonumber
\end{gather*}

\begin{figure}[!ht]
\begin{center}
   \includegraphics[width=0.48\textwidth]{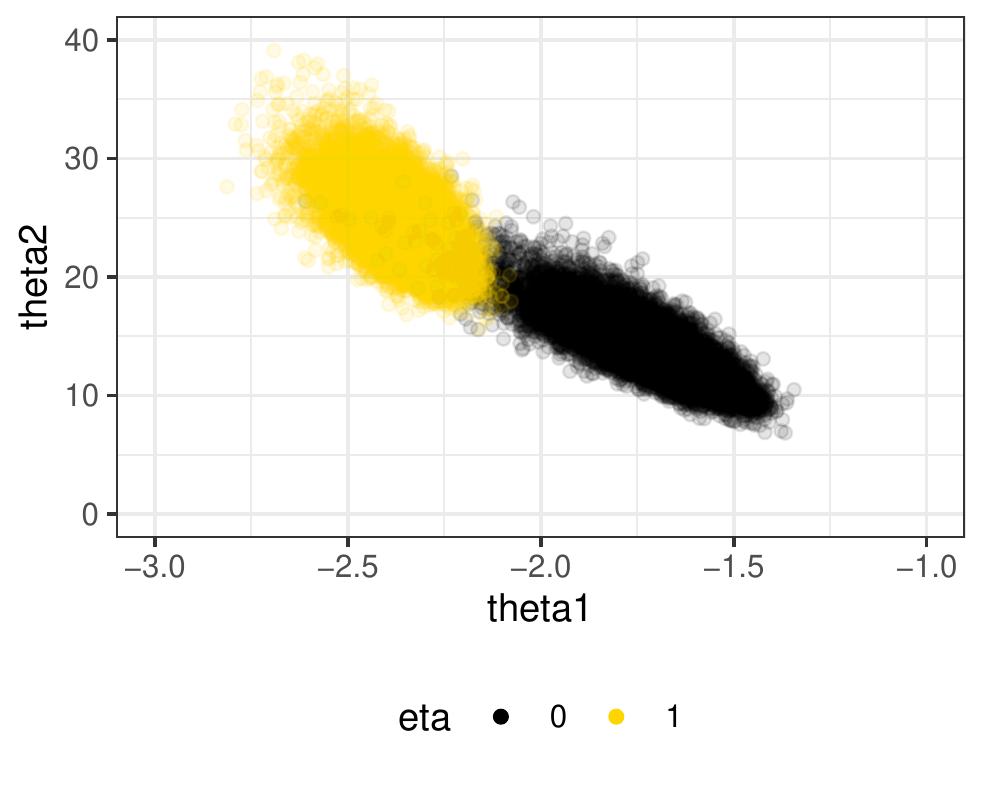}
   \includegraphics[width=0.48\textwidth]{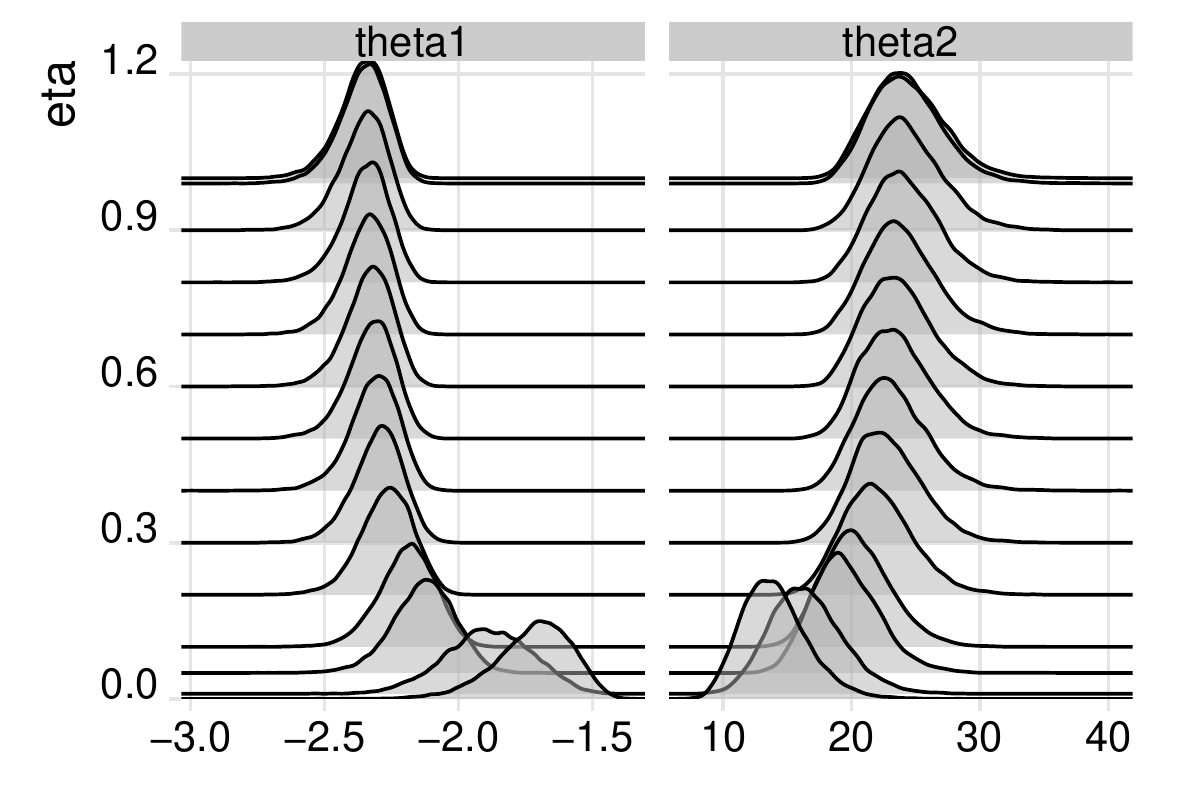}
\end{center}
   \caption{ Joint SMI posterior for $\theta_1$ and $\theta_2$ for the HPV model using MCMC on the SMI posterior with $\eta \in [0,1]$}
   \label{fig:HPV_joint_theta}
\end{figure}

In Fig.~\ref{fig:HPV_joint_theta} we show the posterior distribution for the parameters $\theta_1,\theta_2$. The top panel shows posterior samples for the cut model posterior ($\eta=0$) in black and the full posterior ($\eta=1$) in yellow. The graph agrees with an equivalent plot appearing in \cite{Jacob2017b}. SMI interpolates between these two distributions. The two panels at the bottom of Fig.~\ref{fig:HPV_joint_theta} show the approximate marginal SMI posteriors for the two parameters.

\begin{figure}[!ht]
\begin{center}
   \includegraphics[width=0.48\textwidth]{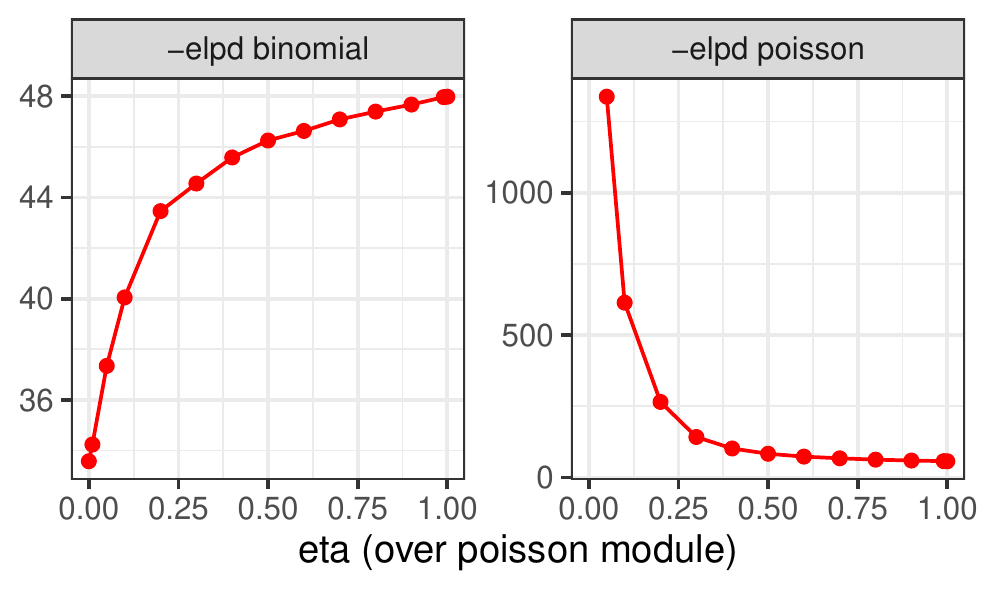}
\end{center}
   \caption{ Estimated elpd (using $WAIC$) as predictive criteria for selection of $\eta\in[0,1]$ for the HPV model. The full model ($\eta=1$) performs better for prediction of the Poisson data, while the cut model ($\eta=0$) dominates for the  Binomial.}
   \label{fig:HPV_elpd_waic}
\end{figure}

Lastly, we follow \cite{Jacob2017b} and evaluate the predictive performance of the various SMI candidate distributions for $\eta \in [0,1]$. Our criterion is the elpd, estimated using the  WAIC, and plotted against $\eta$ in Fig.~\ref{fig:HPV_elpd_waic}. Results expand on but support those reported by \cite{Jacob2017b}: For the task of predicting the Binomial data, the cut model performs best (lower -elpd at $\eta=0$ on the left panel of Fig.~\ref{fig:HPV_elpd_waic}). This is expected as the Poisson module is suspected of being misspecified. Eliminating contamination improves prediction of $Z$. For the task of predicting the Poisson module, full Bayesian analysis performs best (lower -elpd at $\eta=1$ in the right panel of Fig.~\ref{fig:HPV_elpd_waic}) as the information contained in the Binomial data is reliable and helps correct the misspecified module.

\section{Discussion} \label{sec:discussion}

We have given an extension of Bayesian inference to a family of inference procedures indexed by an influence parameter. Our inference procedures bring together Bayesian inference and two qualitatively different inference schemes, power-posteriors and Modular inference/Cut-models, used to treat model misspecification. We show the new family is coherent and falls within the larger loss-based inference framework of \cite{Bissiri2016}.

We gave a straightforward procedure for choosing the inference scheme according to an external $elpd$ criterion, which we implemented using the WAIC and LOOCV. In different examples this selects Bayesian inference, Cut-model inference and interpolating candidate distributions.

When we encounter model misspecification we may consider model elaboration to improve the fit, but we may alternatively expand the inference framework.




\clearpage
\newpage

\bibliography{references} 

\end{document}


\maketitle

In this document we present additional material which complements the main article. The section numbers in this document are aligned with the sections in the main article, for the ease of referencing.

\textbf{Notation: } For the sake of compactness in our derivations, the following expressions are used:
\begin{align*}
p(Z) &= \int p(Z \mid \varphi) \; p(\varphi) \; d\varphi\\
p(Z,Y)_{\theta} &= \int \int p(Z \mid \varphi) \; p( Y \mid \varphi, \theta ) \; p(\varphi,\theta) \; d\varphi \; d\theta\\
p(Z,Y_{\eta})_{\tilde\theta} &= \int \int p(Z \mid \varphi) \; p( Y \mid \varphi, \tilde \theta )^\eta \; p(\varphi,\tilde\theta) \; d\varphi \; d\tilde\theta\\
p(Y_{\eta})_{\tilde\theta} &= \int \int p( Y \mid \varphi, \tilde \theta )^\eta \; p(\varphi,\tilde\theta) \; d\varphi \; d\tilde\theta\\
p(Y,\varphi)_{\theta} &= \int  p( Y \mid \varphi,  \theta ) \; p(\varphi,\theta) \;  d\theta \\
p(Y  \mid  \varphi)_{\theta} &= \frac{1}{p(\varphi)} \; p(Y , \varphi)_{\theta}\\
\end{align*}

All figures and numerical results presented in the main text and this supplement can be replicated using our R package, \texttt{aistats2020smi}, available in Github.
\begin{Schunk}
\begin{Sinput}
> # devtools::install_github("christianu7/aistats2020smi")
> library(aistats2020smi)
\end{Sinput}
\end{Schunk}

\section{Introduction}

Here are a few remarks on model misspecification. Classically, misspecification is identified in goodness-of-fit checks as poor posterior predictive performance on held out data. In this paper, a model is relatively more misspecified if it has relatively worse performance in posterior predictive checks. Notice that this may be caused by a misspecified observation model, as in the HPV example in the main text, but unrepresentative prior assumptions may also lead to mispecification. This is illustrated in Section~\ref{sec:biased_data}. The example in Section~\ref{sec:agric_analysis} arguably suffers from both forms of mispecification.

Like the power posterior (\cite{Walker2001, Bissiri2016, Holmes2017, Grunwald2017, Miller2018a}), the SMI procedure can alternatively be thought of as measuring misspecification. We can measure a model's goodness of fit using the $\eta^*$-values of its modules. Using the degree of influence as a measure of misfit has the advantage that it is defined on a standard scale $[0,1]$ with $\eta^*=1$ corresponding to no evidence for misfit and $\eta^*=0$ indicating the module should be removed entirely when we estimate parameters in the other modules, and indicating substantial misspecification.

\section{Background methods}\label{sec:background}

\subsection{Modular Inference: cut model}
\subsection*{Explicit formulae for cut posterior}

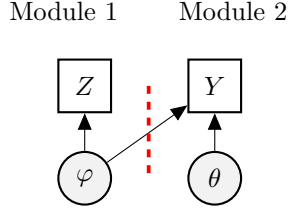
\begin{figure}[!ht]
  \begin{center}
  \begin{tikzpicture}
      \node (Z) [data] {$Z$};
      \node (Y) [data, right=of Z, xshift=0.5cm] {$Y$};
      \node (phi) [param, below=of Z] {$\varphi$};
      \node (theta) [param, below=of Y] {$\theta$};
      \edge {phi} {Z,Y};
      \edge {theta} {Y};
      \draw[dashed,red,line width=0.5mm] (0.85,0) to (0.85,-1.25);
      \node[text width=3cm] at (0.5,1) {Module 1};
      \node[text width=3cm] at (2.75,1) {Module 2};
  \end{tikzpicture}
  \end{center}
  \caption{Graphical representation of a simple multi-modular model.}
  \label{fig:toy_multimodular_model}
\end{figure}

The graphical model analysed in the main text is shown in Figure~\ref{fig:toy_multimodular_model}.

The \textbf{conventional (full) posterior} for this model is:
\begin{align} \label{eq:full_post}
 p(\varphi,\theta \mid Z,Y) &= p(\varphi \mid Z, Y) \; p(\theta \mid Y,\varphi) \nonumber \\
 &= p(Z,Y,\varphi,\theta ) \; \frac{ 1 }{ p(Z,Y)_{\theta} } \nonumber \\
 &= p(Z \mid \varphi) \; p(Y \mid \varphi, \theta ) \; p(\varphi, \theta) \; \frac{ 1 }{ p(Z,Y)_{\theta} }
\end{align}

The \textbf{cut posterior} for this model is defined \citep{Plummer2015} as:
\begin{align} \label{eq:cut_post}
 p_{cut}(\varphi,\theta \mid Z,Y) &= p(\varphi \mid Z) \; p(\theta \mid Y,\varphi) \nonumber \\
 &= \frac{ p(Z \mid \varphi) p(\varphi) }{ p(Z) } \frac{ p( Y \mid \varphi,  \theta ) \; p(\varphi,\theta) }{ p( Y,\varphi)_{\theta}} \nonumber \\
 &= p(Z \mid \varphi) \; p(Y \mid \varphi, \theta ) \; \frac{ 1 }{ p(Z)} \; \frac{ 1 }{p(Y \mid \varphi)_{\theta} }  p(\varphi, \theta)
\end{align}

Note the relation between the cut posterior and the conventional posterior
\begin{align*}
 p_{cut}(\varphi,\theta \mid Z,Y) &= p(Z, Y , \varphi, \theta ) \; \frac{ 1 }{ p(Z) \; p(Y \mid \varphi)_{\theta} } \\
 &= p(\varphi, \theta \mid Z, Y ) \; \frac{ p(Z, Y) }{ p(Z) \; p(Y \mid \varphi)_{\theta} } \\
\end{align*}

\section{Semi-Modular Inference} \label{sec:smi}

\subsection{SMI distributions}
\subsection*{Explicit formulae for SMI posterior}

The \textbf{$\eta$-smi posterior} is defined as
\begin{align} \label{eq:smi_post}
 p_{smi,\eta}(\varphi,\theta,\tilde\theta \mid Z, Y ) &= p_{pow,\eta}(\varphi,\tilde\theta \mid Z,Y) \; p(\theta \mid Y,\varphi) \nonumber \\
 &= \frac{ p(Z \mid \varphi) \; p( Y \mid \varphi, \tilde \theta )^\eta \; p(\varphi,\tilde\theta) }{ p(Z, Y_{\eta})_{\tilde\theta}} \frac{ p( Y \mid \varphi,  \theta ) \; p(\varphi,\theta) }{ p( Y , \varphi)_{\theta} } \nonumber \\
 &= p(Z \mid \varphi) \; p(Y \mid \varphi, \tilde\theta )^{\eta} p(Y \mid \varphi, \theta ) \; \frac{1}{ p(Z, Y_{\eta} )_{\tilde\theta} \; p(Y \mid \varphi)_{\theta} } p(\varphi, \theta, \tilde\theta) \\
 &\propto p(Z \mid \varphi) \; p(Y \mid \varphi, \tilde\theta )^{\eta} p(Y \mid \varphi, \theta ) \; \frac{1}{ p(Y \mid \varphi)_{\theta} } p(\varphi, \theta, \tilde\theta) \nonumber
\end{align}

In the penultimate step, we assume that $\theta$ and $\tilde\theta$ are conditionally independent given $\varphi$ in the prior, so $ p(\varphi, \theta, \tilde\theta) = p(\theta \mid \varphi) p(\tilde\theta\mid \varphi) p(\varphi)= p( \varphi, \theta ) \; p( \varphi, \tilde\theta ) \frac{1}{p(\varphi)} $.

From here is easy to see two particular cases of the $\eta$-smi posterior: taking marginals over $\tilde\theta$ in SMI, the cut model $p_{smi,\eta}(\varphi,\theta \mid Z,Y) = p_{cut}(\varphi,\theta \mid Z,Y)$ when $\eta=0$; and the conventional posterior
 $p_{smi,\eta}(\varphi,\theta \mid Z,Y) = p(\varphi,\theta \mid Z,Y)$ when $\eta=1$.

\section{Analysis with (Semi-)Modular Inference} \label{sec:smi_considerations}

\subsection{Coherence of (Semi-)Modular Inference}

\subsubsection{Coherent update of beliefs} \label{sec:bissiri_summary}
In \cite{Bissiri2016}, the conventional update of beliefs provided by Bayes theorem is expanded, providing a generalised framework in which alternative inferential options are justified beyond the conventional posterior.

Let's use single-module notation in this Subsection~\ref{sec:bissiri_summary}, so we are aligned with \cite{Bissiri2016}. Denote newly observe data as $Y$ and the parameter of interest $\theta$.

The framework established by \cite{Bissiri2016} relies on the idea that an update of beliefs must exist. This update of beliefs is performed, under a decision theory framework, by a function $\psi$, which turns the prior into posterior beliefs by incorporating new observed data $y$ via a loss function $l(\theta;y)$. that is:
\begin{equation*}
  p(\theta \mid y) = \psi\{ l(\theta;y) , p(\theta) \}
\end{equation*}

Such update $\psi$ is \textbf{Coherent} if it ensures that we preserve the posterior regardless of the order in which the data was observed. In other words, the posterior is the same whether we update our beliefs by observing all data simultaneously or by observing the data sequentially,
\begin{equation*}
  \psi\{ l(\theta;x_2) , \psi\{ l(\theta;x_1) , p(\theta) \} \} = \psi\{ l(\theta;x_1)+l(\theta;x_2) , p(\theta) \}
\end{equation*}

The authors show that an optimal, valid and coherent update of beliefs is of the form
\begin{equation*}
	p(\theta \mid y ) = \psi\{ l(\theta;y) , p(\theta) \} = \frac{ \exp\{ -l(\theta;y) \} \; p(\theta) }{\int \exp\{ -l(\theta;y) \} \; p(\theta) \; d\theta}
\end{equation*}

\subsubsection{Coherence of the SMI posterior} \label{sec:suppl_cut_consistency}

Back in our multi-modular setting, the flexibility of the framework allows us to analyse the SMI posterior and the cut model posterior as valid schemes for the update of beliefs, which differ from the traditional fully-Bayesian update.

In the proofs which follow we work with SMI at an arbitrary fixed $\eta$. Since this means the results hold at $\eta=0$, they hold for modular inference/cut models/Bayesian multiple imputation, at least for imputation in the modular setting we consider.

From Eq.~\ref{eq:cut_post} we can see that the loss function underlying the update of beliefs in the cut model is
\begin{equation} \label{eq:cut_loss}
  l_{cut}( (\varphi,\theta) ; (Z,Y) ) = - \log p(Z \mid \varphi) -\log p(Y \mid \varphi,\theta) + \log p(Y \mid \varphi).
\end{equation}

Similarly, from Eq.~\ref{eq:smi_post} we can derive the loss function underlying the update of beliefs in SMI the posterior,
\begin{equation} \label{eq:smi_loss}
  l_{smi,\eta}( (\varphi,\theta,\tilde\theta) ; (Z,Y) ) = - \log p(Z \mid \varphi) - \eta \log p(Y \mid \varphi,\tilde\theta) -\log p(Y \mid \varphi,\theta) + \log p(Y \mid \varphi) .
\end{equation}

Here we prove that the SMI posterior preserves multi-modular coherence. In multi-modular settings, coherence must hold in two ways: 1) by observing responses from different modules one after the other (i.e. first $Z$, and then $Y$); and 2) by observing sequential fragments within the same module (e.g. first $Z_1$, and then $Z_2$, with $Z=(Z_1,Z_2)$).

The SMI posterior in Eq.~\ref{eq:smi_post} updates the belief distribution by observing the two datasets simultaneously. Under coherent inference, we can also update by observing data only from one module at a time, and still preserve the loss function in eq.\ref{eq:smi_loss}.

Say our current distribution of beliefs about $(\varphi,\theta, \tilde\theta)$ is $p(\varphi,\theta, \tilde\theta)$. Our updated belief by observing \textit{only} $Z$ would be
\begin{align} \label{eq:smi_post_z}
 p_{smi,\eta}(\varphi,\theta,\tilde\theta \mid Z) &= \psi\{ l( (\varphi,\theta,\tilde\theta); Z ) , p( \varphi, \theta,\tilde\theta) \} \nonumber \\
 &= p(Z \mid \varphi) \; \frac{1}{ p(Z ) } p(\varphi, \theta, \tilde\theta) \\
 &= p(Z \mid \varphi) \frac{ 1 }{ \int p(Z \mid \varphi) \; p(\varphi) \; d\varphi } p(\varphi, \theta, \tilde\theta) \nonumber
\end{align}

similarly, if we observe \textit{only} $Y$ our updated beliefs are
\begin{align} \label{eq:smi_post_y}
 p_{smi,\eta}(\varphi,\theta,\tilde\theta \mid Y) &= \psi\{ l( (\varphi,\theta,\tilde\theta); Y ) , p( \varphi, \theta,\tilde\theta) \} \nonumber \\
 &= p(Y \mid \varphi, \tilde\theta )^{\eta} p(Y \mid \varphi, \theta ) \; \frac{1}{ p( Y_{\eta} )_{\tilde\theta} \; p(Y \mid \varphi)_{\theta} } p(\varphi, \theta, \tilde\theta) \\
 &= p( Y \mid \varphi, \tilde \theta )^\eta \; p( Y \mid \varphi,  \theta ) \frac{ 1 }{ \int \int p( Y \mid \varphi, \tilde \theta )^\eta \; p(\varphi,\tilde\theta) \; d\varphi \; d\tilde\theta} \frac{ p(\varphi) }{ \int  p( Y \mid \varphi,  \theta ) \; p(\varphi, \theta ) \;  d\theta} p(\varphi, \theta, \tilde\theta) \nonumber
\end{align}

\subsubsection*{Coherence when observing data for different modules sequentially.}

First, we show that the update from prior to posterior, is equivalent to updating sequentially first $Z$ and then $Y$. This is (a)=(b)+(c) in the following diagram:

\begin{tikzcd}
p(\varphi,\theta,\tilde\theta) \ar[rr,"(a)", bend left=10] \ar[r,"(b)", swap] & p_{smi,\eta}(\varphi,\theta,\tilde\theta \mid Z) \ar[r,"(c)", swap] & p_{smi,\eta}(\varphi,\theta,\tilde\theta \mid Z,Y)
\end{tikzcd}

The update (a) is Eq.~\ref{eq:smi_post}, updating beliefs by only observing both $Z$ and $Y$
\begin{equation*}
    p_{(a)}(\varphi,\theta,\tilde\theta \mid Z, Y) = p_{smi,\eta}(\varphi,\theta,\tilde\theta \mid Z, Y)
\end{equation*}

The update (b) is similar to Eq.~\ref{eq:smi_post_z}, updating beliefs only with data $Z$.
\begin{align} \label{eq:smi_update_b}
 p_{(b)}(\varphi,\theta,\tilde\theta \mid Z) &= p_{smi,\eta}(\varphi,\theta,\tilde\theta \mid Z) \\
 &= p(Z \mid \varphi) \; \frac{1}{ p(Z)} p(\varphi, \theta, \tilde\theta) \nonumber \\
 &= p(Z \mid \varphi) \; \frac{1}{  \int p(Z \mid \varphi) \; p(\varphi) \; d\varphi } p(\varphi, \theta, \tilde\theta).
\end{align}

The update (c) is equivalent to Eq.~\ref{eq:smi_post_y} substituting the current beliefs with $p_{smi,\eta}(\varphi,\theta,\tilde\theta \mid Z)$, and updating beliefs only with data $Y$.
\begin{align*}
 p_{(b)+(c)}(\varphi,\theta,\tilde\theta \mid Z, Y) &\propto p( Y \mid \varphi, \tilde \theta )^\eta p( Y \mid \varphi,  \theta ) \frac{ p_{(b)}(\varphi \mid Z) }{ \int  p( Y \mid \varphi,  \theta ) \; p_{(b)}(\varphi, \theta \mid Z ) \;  d\theta} p_{(b)}(\varphi, \theta, \tilde\theta \mid Z)\\
 &\propto p(Z \mid \varphi) p( Y \mid \varphi, \tilde \theta )^\eta p( Y \mid \varphi,  \theta ) \frac{ 1 }{ P(Y \mid \varphi )_{\theta} } p(\varphi, \theta, \tilde\theta).
\end{align*}

The equivalence $p_{(b)+(c)}(\varphi,\theta,\tilde\theta \mid Z, Y)=p_{smi,\eta}(\varphi,\theta,\tilde\theta \mid Z,Y)$ is clear by comparing the last formula with smi posterior in Eq.~\ref{eq:smi_post}. For the last line, we used the following identity
\begin{align*}
 \frac{ p_{(b)}(\varphi \mid Z) }{ \int  p( Y \mid \varphi,  \theta ) \; p_{(b)}(\varphi, \theta \mid Z ) \;  d\theta } &=\frac{ \int \int p_{(b)}(\varphi, \theta, \tilde\theta\mid Z) d\theta d\tilde\theta }{ \int  \int p( Y \mid \varphi,  \theta ) \; p_{(b)}(\varphi, \theta, \tilde\theta\mid Z ) \;  d\theta d\tilde\theta } \\
 &= \frac{ \int \int p(Z \mid \varphi) \; \frac{1}{  \int p(Z \mid \varphi) \; p(\varphi) \; d\varphi } p(\varphi, \theta, \tilde\theta) d\theta d\tilde\theta}{ \int \int p( Y \mid \varphi,  \theta ) \; p(Z \mid \varphi) \; \frac{1}{  \int p(Z \mid \varphi) \; p(\varphi) \; d\varphi } p(\varphi, \theta, \tilde\theta) \;  d\theta d\tilde\theta } \\
 &=\frac{ p(\varphi ) }{ \int  p( Y \mid \varphi,  \theta ) \; p(\varphi,\theta) \;  d\theta} \\
 &= \frac{1}{p(Y \mid \varphi)_{\theta}}.
\end{align*}

Now, we want a similar result by first observing $Y$ and then $Z$, i.e. (a)=(d)+(e) in the following diagram

\begin{tikzcd}
p(\varphi,\theta,\tilde\theta) \ar[rr,"(a)", bend left=10] \ar[r,"(d)", swap] & p_{smi,\eta}(\varphi,\theta,\tilde\theta \mid Y) \ar[r,"(e)", swap] & p_{smi,\eta}(\varphi,\theta,\tilde\theta \mid Z,Y)
\end{tikzcd}

The update (a) is again given by Equation~\ref{eq:smi_post}.

The update (d) is Eq.~\ref{eq:smi_post_y}
\begin{equation} \label{eq:smi_update_d}
 p_{(d)}(\varphi,\theta,\tilde\theta \mid Y) \propto p(Y \mid \varphi, \tilde\theta )^{\eta} p(Y \mid \varphi, \theta ) \; \frac{1}{ p(Y \mid \varphi)_{\theta} } p(\varphi, \theta, \tilde\theta).
\end{equation}

The update (d)+(e) is equivalent to Eq.~\ref{eq:smi_post_z} substituting the current beliefs with $p_{(d)}(\varphi,\theta,\tilde\theta \mid Y)$
\begin{align*}
 p_{(d)+(e)}(\varphi,\theta,\tilde\theta \mid Z, Y) &\propto p( Z \mid \varphi) \; p_{(d)}(\varphi, \theta, \tilde\theta \mid Y ) \\
 &= p( Z \mid \varphi) \; p(Y \mid \varphi, \tilde\theta )^{\eta} \; p(Y \mid \varphi, \theta ) \; \frac{1}{ p(Y \mid \varphi)_{\theta} } p(\varphi, \theta, \tilde\theta).
\end{align*}
The equivalence $p_{(d)+(e)}(\varphi,\theta,\tilde\theta \mid Z, Y)=p_{smi,\eta}(\varphi,\theta,\tilde\theta \mid Z,Y)$ is direct from comparing the last formula with Eq.~\ref{eq:smi_post}.

\subsubsection*{Coherence when observing data partitioned from the same module.}

We now verify that the SMI posterior is coherent when observing a sequential portions of the same module. Define the partitions $Z=(Z_1,Z_2)$ and $Y=(Y_1,Y_2)$.

First, we verify coherence for the partition of data $Z$. We want to check (b)=(b1)+(b2) in the following diagram.
\begin{tikzcd}
p(\varphi,\theta,\tilde\theta) \ar[rr,"(b)", bend left=10] \ar[r,"(b1)", swap] & p_{smi,\eta}(\varphi,\theta,\tilde\theta \mid Z_1) \ar[r,"(b2)", swap] & p_{smi,\eta}(\varphi,\theta,\tilde\theta \mid Z)
\end{tikzcd}

Update (b) is the same as defined above in Equation~\ref{eq:smi_update_b}

Updates (b1) and (b2) are similar to Eq.~\ref{eq:smi_post_z}, substituting the corresponding $Z$ and current state of beliefs
\begin{align*}
 p_{(b1)}(\varphi,\theta,\tilde\theta \mid Z_1) &= p(Z_1 \mid \varphi) \; \frac{1}{  \int p(Z_1 \mid \varphi) \; p(\varphi) \; d\varphi } p(\varphi, \theta, \tilde\theta), \\
 p_{(b1)+(b2)}(\varphi,\theta,\tilde\theta \mid Z_1,Z_2) &= p(Z_2 \mid \varphi) \; \frac{1}{  \int p_{(b1)}(Z_2 \mid \varphi) \; p_{(b1)}(\varphi \mid Z_1) \; d\varphi } p_{(b1)}(\varphi, \theta, \tilde\theta \mid Z_1) \\
 &\propto p(Z_1 \mid \varphi) \; p(Z_2 \mid \varphi) \; p(\varphi, \theta, \tilde\theta ) \\
 &= p(Z \mid \varphi) \; p(\varphi, \theta, \tilde\theta ),
\end{align*}
clearly $p_{(b1)+(b2)}(\varphi,\theta,\tilde\theta \mid Z_1,Z_2)=p_{(b)}(\varphi,\theta,\tilde\theta \mid Z)$.

Lastly, we verify coherence for the partition of data $Y$. We want to check (d)=(d1)+(d2) in the following diagram.

\begin{tikzcd}
p(\varphi,\theta,\tilde\theta) \ar[rr,"(d)", bend left=10] \ar[r,"(d1)", swap] & p_{smi,\eta}(\varphi,\theta,\tilde\theta \mid Y_1) \ar[r,"(d2)", swap] & p_{smi,\eta}(\varphi,\theta,\tilde\theta \mid Y)
\end{tikzcd}

Update (d) is the same as defined above in Equation~\ref{eq:smi_update_d}

Updates (d1) and (d2) are similar to Eq.~\ref{eq:smi_post_y}, substituting the corresponding $Y$ and current state of beliefs
\begin{align*}
 p_{(d1)}(\varphi,\theta,\tilde\theta \mid Y_1) = & p(Y_1 \mid \varphi, \tilde\theta )^{\eta} p(Y_1 \mid \varphi, \theta ) \; \frac{1}{ p( Y_{1 ,\eta} )_{\tilde\theta} \; p(Y_1 \mid \varphi)_{\theta} } p(\varphi, \theta, \tilde\theta) \\
 \propto & p(Y_1 \mid \varphi, \tilde\theta )^{\eta} p(Y_1 \mid \varphi, \theta ) \; \frac{1}{ p(Y_1 \mid \varphi)_{\theta} } p(\varphi, \theta, \tilde\theta) \\
 = & p( Y_1 \mid \varphi, \tilde \theta )^\eta \; p( Y_1 \mid \varphi,  \theta ) \frac{ p(\varphi) }{ \int  p( Y_1 \mid \varphi,  \theta ) \; p(\varphi, \theta ) \;  d\theta} p(\varphi, \theta, \tilde\theta) \nonumber \\
 p_{(d1)+(d2)}(\varphi,\theta,\tilde\theta \mid Y_1,Y_2) \propto & p( Y_2 \mid \varphi, \tilde \theta )^\eta \; p( Y_2 \mid \varphi,  \theta ) \frac{ p_{(d1)}(\varphi) }{ \int  p( Y_2 \mid \varphi,  \theta ) \; p_{(d1)}(\varphi, \theta ) \;  d\theta} p_{(d1)}(\varphi, \theta, \tilde\theta) \nonumber \\
\propto & \left( p(Y_1 \mid \varphi, \tilde\theta ) \; p(Y_2 \mid \varphi, \tilde\theta ) \right)^{\eta} \left( p(Y_1 \mid \varphi, \theta ) \; p(Y_2 \mid \varphi, \theta ) \right) \cdot \\
 & \cdot \frac{ p_{(d1)}(\varphi) }{ \int  p( Y_2 \mid \varphi,  \theta ) \; p_{(d1)}(\varphi, \theta ) \;  d\theta} \frac{ 1 }{ p( Y_1 \mid \varphi )_{\theta} } p(\varphi,\theta,\tilde\theta) \\
\propto & p(Y \mid \varphi, \tilde\theta )^{\eta} p(Y \mid \varphi, \theta ) \; \frac{1}{ p(Y \mid \varphi)_{\theta} } p(\varphi, \theta, \tilde\theta).
\end{align*}
from here is clear that $p_{(d1)+(d2)}(\varphi,\theta,\tilde\theta \mid Y_1,Y_2)=p_{(d)}(\varphi,\theta,\tilde\theta \mid Y)$. In the last step, we used the following identity
\begin{align*}
\frac{ p_{(d1)}(\varphi) }{ \int  p( Y_2 \mid \varphi,  \theta ) \; p_{(d1)}(\varphi, \theta ) \;  d\theta}  &= \frac{ \int \int p_{(d1)}(\varphi, \theta, \tilde\theta) d\theta d\tilde\theta }{ \int \int p( Y_2 \mid \varphi,  \theta ) \; p_{(d1)}(\varphi, \theta, \tilde\theta ) \;  d\theta d\tilde\theta}\\
&\propto \frac{ \int \int p( Y_1 \mid \varphi, \tilde \theta )^\eta \; p( Y_1 \mid \varphi,  \theta ) \frac{1}{ p(Y_1 \mid \varphi)_{\theta} } p(\varphi, \theta, \tilde\theta) d\theta d\tilde\theta }{ \int \int p( Y_2 \mid \varphi,  \theta ) \; p( Y_1 \mid \varphi, \tilde \theta )^\eta \; p( Y_1 \mid \varphi,  \theta ) \frac{1}{ p(Y_1 \mid \varphi)_{\theta} } p(\varphi, \theta, \tilde\theta) \;  d\theta d\tilde\theta} \\
&= \frac{ \frac{1}{p(\varphi)} \left( \int p( Y_1 \mid \varphi, \tilde \theta )^\eta p(\varphi, \tilde\theta) d\tilde\theta \right) \left(\int  p( Y_1 \mid \varphi,  \theta ) p(\varphi, \theta) d\theta \right) }{ \frac{1}{p(\varphi)} \left( \int  p( Y_1 \mid \varphi, \tilde \theta )^\eta p(\varphi, \tilde\theta) \;  d\tilde\theta \right) \left(  \int p( Y_1 \mid \varphi,  \theta ) p( Y_2 \mid \varphi,  \theta ) \; p(\varphi, \theta ) \;  d\theta \right) }\\
&=\frac{ p(Y_1 \mid \varphi)_{\theta} }{ p(Y \mid \varphi)_{\theta} }
\end{align*}
here again, we assumed that $\theta$ and $\tilde\theta$ are conditionally independent given $\varphi$ in the prior, so $ p(\varphi, \theta, \tilde\theta) = p( \varphi, \theta ) \; p( \varphi, \tilde\theta ) \frac{1}{p(\varphi)} $.

\subsection{Targeting the modular posterior}

\subsubsection{Detailed balance of SMI posterior}

Here we show that the $\eta$-smi posterior (and cut posterior in particular) preserves the detailed balance condition when we use the transition kernel implied by the two-stage MCMC algorithm proposed in the main text.

\begin{enumerate}
    \item Sample $(\varphi,\tilde\theta) \sim p_{pow,\eta}(\varphi, \tilde\theta \mid Z, Y) = p(Z \mid \varphi) \; p(Y \mid \varphi, \tilde\theta )^{\eta} \frac{1}{ p(Z, Y_{\eta} )_{\tilde\theta} } p(\varphi, \tilde\theta) $
    \item Sample $\theta \sim p( \theta \mid Y,\varphi )=p(Y \mid \varphi, \theta ) \frac{1}{ p(Y \mid \varphi)_{\theta} } p(\theta) $
\end{enumerate}

The first step updates $(\varphi,\tilde\theta)$ using the powered likelihood. It is not difficult to target this posterior using traditional sampling methods.

The second term updates $\theta$ exactly from its conditional posterior given data $Y$ from module 2, and a \textit{fixed} value $\varphi$.

The transition kernel for one iteration in this scheme is given by
\begin{equation} \label{eq:smi_kernel}
K(\varphi',\theta',\tilde\theta' \mid \varphi,\theta,\tilde\theta)= K(\varphi',\tilde\theta' \mid \varphi,\tilde\theta) p(\theta' \mid Y,\varphi').
\end{equation}

By construction, the first stage of the update (using the powered likelihood) is in detailed balance with the powered likelihood, i.e. satisfies
\begin{equation*}
p_{pow,\eta}(\varphi,\tilde\theta \mid Z,Y) K(\varphi',\tilde\theta' \mid \varphi,\tilde\theta) = p_{pow,\eta}(\varphi',\tilde\theta' \mid Z,Y) K(\varphi,\tilde\theta \mid \varphi',\tilde\theta')
\end{equation*}

From here we see that the $\eta$-smi posterior in Eq.~\ref{eq:smi_post} satisfies detailed balance with the transition kernel in eq.
\ref{eq:smi_kernel}

\begin{align*}
p_{smi,\eta}&(\varphi,\theta,\tilde\theta \mid Z,Y) K(\varphi',\theta',\tilde\theta' \mid \varphi,\theta,\tilde\theta) \\
&= [ p_{pow,\eta}(\varphi, \tilde\theta \mid Z,Y)p(\theta \mid Y,\varphi) ][ K(\varphi',\tilde\theta' \mid \varphi,\tilde\theta) p(\theta' \mid Y,\varphi') ] \\
&= p_{pow,\eta}(\varphi, \tilde\theta \mid Z,Y) p(\theta \mid Y,\varphi) \frac{p_{pow,\eta}(\varphi',\tilde\theta' \mid Z,Y) K(\varphi,\tilde\theta \mid \varphi',\tilde\theta')}{p_{pow,\eta}(\varphi,\tilde\theta \mid Z,Y)} p(\theta' \mid Y,\varphi')\\
&= p_{pow,\eta}(\varphi',\tilde\theta' \mid Z,Y) p(\theta' \mid Y,\varphi') K(\varphi,\tilde\theta \mid \varphi',\tilde\theta') p(\theta \mid Y,\varphi)\\
&=p_{smi,\eta}(\varphi',\theta',\tilde\theta' \mid Z,Y) K(\varphi,\theta,\tilde\theta \mid \varphi',\theta',\tilde\theta')
\end{align*}

\subsubsection{Further details about Nested MCMC for SMI posterior.}

Our implementation of MCMC targeting the SMI posterior is a two-stage sampler described in Algorithm 1 in the main article.

This is arguably the simpler approach that has been discussed in literature about MCMC targeting a modular posterior (e.g. Unbiased MCMC via couplings \citep{Jacob2017}).

Algorithm 1 does not make specific assumptions about the class of MCMC sampler that is used at each stage. The only requirement is to check convergence in both stages. For the first step, sampling $\phi$, we proceed as any traditional implementation of MCMC, running the chains until we have satisfied classical convergence test. The second step, sampling $\theta$, we only need to guarantee that the last sample in every sub-chain are taken after we reached the equilibrium distribution.

Our examples in Section~\ref{sec:data_analyses} use standard MCMC at each step. For the agricultural data in Sec.\ref{sec:agric_analysis} we use random walk Metropolis-Hastings in both stages. For the epidemiological data, the samplers for both stages are implemented in Stan \citep{Stan2017}, using Hamiltonian Monte Carlo. In both cases we performed convergence analysis for the main Chain (step 1), and experimented with various lengths for the sub-chain, in the end, we choose 500 iterations as a conservative length that guaranteed the last iteration was sampled from the equilibrium distribution. The detailed implementation can be found in the accompanying R package \texttt{aistats2020smi} available in GitHub \footnote{\url{https://github.com/christianu7/aistats2020smi}}.

\setcounter{subsection}{3}
\subsection{Computational cost of SMI}

Our baseline is standard Bayes-MCMC on the original full model. We suppose for simplicity this was implemented using separate updates for $\theta|\varphi$ and $\varphi|\theta$, though these need not be Gibbs updates.
Let $\tau_{\varphi,\theta}$ be the Integrated Autocorrelation Time (IACT) of Bayes-MCMC. If the Effective Sample Size (ESS) of the full Bayes-MCMC output is $N$ then we must have done $T=N\tau_{\varphi,\theta}$ MCMC steps. If {\it one Bayes-MCMC step updating both $\theta$ and $\varphi$ has unit cost} then the overall cost is $W_{bm}=T$[time]. This doesn't parallelise.

The SMI posterior is \[
p_{\eta-smi}(\varphi,\tilde\theta,\theta|Y,Z)=p_{pow,\eta}(\varphi,\tilde\theta|Y,Z)p(\theta|Y,\varphi).
\]
If we use the same Bayes-MCMC updates to sample $p_{pow,\eta}(\varphi,\tilde\theta|Y,Z)$ then the work sampling $(\varphi,\tilde\theta)$ at one $\eta$-value is $W_{Bayes}$. Thin the $\varphi$ samples every $\tau_{\varphi,\theta}$ steps to get an ESS about $N$ (this is rough, because the target changes with $\eta$, but reasonable if we allow for some tuning of the MCMC with $\eta$ - we didnt need to tune in our examples).

We use the Bayes-MCMC update for $p(\theta|Y,\varphi)$ in SMI. Let $\tau_\theta$ be the IACT. Typically, $\tau_\theta<\tau_{\varphi,\theta}$ (the target has lower dimension; illustrative proofs can be given in simple special cases) so take $\tau_\theta=\tau_{\varphi,\theta}$ (conservative, and note that these ``side-chains'' targeting $p(\theta|Y,\varphi)$ can be started close to equilibrium using the $\theta$-value output from sampling at the previous step).
We run the $\theta$-sampler to equilibrium (initialise $\theta^{(t)}$-run with $\theta^{(t-1)}$). Suppose this takes $K\tau_\theta$ steps ($K\approx 5$ is reasonable).

In the following we assume simulation at different $\eta$-values is parallelized over machines, while simulation of $\theta|\phi$ is parallelized over threads on a machine. Suppose we have $M_{t}$ threads on each of $M_{p}$ machines. The $\theta$-sampling parallelizes with a small communication overhead if the time to do $K\tau_{\varphi,\theta}$ of the $\theta|\varphi$-updates is significantly larger than the communication time.
The cost of the $\theta$-update in the second stage of SMI is no more than the cost of one update in the original Bayes-MCMC where both $\theta$ and $\varphi$ were updated, so a runtime cost for $\theta$-updates equal one unit is conservative. It follows that $(\tilde\theta,\varphi,\theta)$-sampling at one $\eta$-value costs about \[
W_{\eta}=W_{bm}(1+K/M_{t}).
\]

We have to repeat this $J$ times, sampling $p_{\eta-smi}(\varphi,\tilde\theta,\theta|Y,Z)$ for each of $J$ different $\eta$-values spaced between $\eta=0$ and $\eta=1$ ($J\approx 20$ should be enough) using $M_{p}$ machines. This larger task parallelises essentially perfectly. The total cost is
\[
W_{smi}=W_{bm}(1+K/M_{t}) \times J/M_{p}.
\]
For eg if we assign resources as $M_{p}=J$ and $M_{t}=1$ (just parallelise over $\eta$) the SMI cost is not worse than about $10$ times the cost of doing Bayes-MCMC. This reflects our experience.

Finally, we compute and smooth the WAIC across the $J$ runs at different $\eta$-values. This part is fast output-analysis. The ESS must be big enough to get stable WAIC estimates, but WAIC is "nice" to estimate. Very roughly,
\[W_{smi}\simeq 10 W_{bm}\]
should be achievable without a great deal of work on top of the cost of implementing and running standard Bayes-MCMC.

\section{Data Analyses} \label{sec:data_analyses}

\subsection{Simulation study: Biased data} \label{sec:biased_data}

\begin{multicols}{2}
Model:
\begin{align*}
  Z \mid \varphi &\sim N( \varphi, \sigma_z^2 ) \\
  Y \mid \varphi, \theta  &\sim N( \varphi + \theta, \sigma_y^2 )
\end{align*}
with $\sigma_z^2$ and $\sigma_y^2$ (and other $\sigma$'s) known.

Priors:
\begin{align*}
  \varphi &\sim N( 0, \sigma_\varphi^2 ) \\
  \theta &\sim N( 0, \sigma_\theta^2 ) \\
  \tilde\theta &\sim N( 0, \tilde\sigma_\theta^2 )
\end{align*}
\end{multicols}

In our example on the main text, we know the generative parameters: $(\varphi^*,\theta^*, \tilde\theta^*)$. We compare these \textit{true} values with the estimates arising from the $\eta$-smi posterior, for different values of $\eta\in[0,1]$.


\subsubsection{SMI posterior}

First, derive $p(Y \mid \varphi)_{\theta}$ as a function of $\varphi$
\begin{align*}
  p(Y \mid \varphi)_{\theta} &= \frac{1}{p(\varphi)} \; \int  p( Y \mid \varphi,  \theta ) \; p(\varphi,\theta) \;  d\theta \\
  &\propto \exp\{ -\frac{1}{2} [ \varphi^2 ( \frac{m}{m \sigma_\theta^2 + \sigma_y^2 } ) - 2 \varphi (\bar Y \frac{m}{m \sigma_\theta^2 + \sigma_y^2 }) ] \}
\end{align*}

Now we can obtain the $\eta$-smi posterior
\begin{align*}
 p_{smi,\eta}(\varphi,\theta,\tilde\theta \mid Z, Y ) &= p(Z \mid \varphi) \; p(Y \mid \varphi, \tilde\theta )^{\eta} p(Y \mid \varphi, \theta ) \; \frac{1}{ p(Z, Y_{\eta} )_{\tilde\theta} \; p(Y \mid \varphi)_{\theta} } p(\varphi, \theta, \tilde\theta) \\
 &\propto \exp\{ -\frac{1}{2} \; [ \varphi^2 ( \frac{n}{\sigma_z^2} + \frac{m}{\sigma_y^2}(1+\eta) + \frac{1}{\sigma_\varphi^2} - \frac{m}{\sigma_y^2 + m \sigma_\theta^2} ) - 2 \varphi (\frac{n}{\sigma_z^2} + \frac{m}{\sigma_y^2}(1+\eta) - \bar Y \frac{m}{ \sigma_y^2 + m \sigma_\theta^2}) + \\
 & \hspace{2cm} \theta^2 (\frac{m}{\sigma_y^2} + \frac{1}{\sigma_\theta^2}) - 2 \theta (\bar Y \frac{m}{\sigma_y^2 }) + \\
 & \hspace{2cm} \tilde\theta^2 ( \eta \frac{m}{\sigma_y^2} + \frac{1}{\tilde\sigma_\theta^2}) - 2 \tilde\theta (\eta \bar Y \frac{m}{\sigma_y^2 }) + \\
 & \hspace{2cm} + 2 \varphi \theta (\frac{m}{\sigma_y^2}) + 2 \varphi \tilde\theta (\eta \frac{m}{\sigma_y^2 }) ] \}
\end{align*}

From here we see that the joint posterior distribution for $(\varphi,\theta,\tilde\theta)$ is a multivariate normal distribution defined as:
\begin{equation} \label{eq:smi_post_5_1}
p_{smi,\eta}(\varphi,\theta,\tilde\theta \mid Z, Y ) = \text{Normal}( \mu, \Sigma ),
\end{equation}

with
\begin{equation*}
  \Sigma = \begin{bmatrix}
  \frac{n}{\sigma_z^2} + \frac{m}{\sigma_y^2}(1+\eta) - \frac{m}{ \sigma_y^2 + m \sigma_\theta^2 } + \frac{1}{\sigma_\varphi^2} & \frac{m}{\sigma_y^2} & \eta \frac{m}{\sigma_y^2 } \\
  \frac{m}{\sigma_y^2} & \frac{m}{\sigma_y^2} + \frac{1}{\sigma_\theta^2} & 0 \\
  \eta \frac{m}{\sigma_y^2 } & 0 &  \eta \frac{m}{\sigma_y^2} + \frac{1}{\tilde\sigma_\theta^2}
\end{bmatrix}^{-1} \text{, and } \mu = \Sigma \begin{bmatrix}
  \frac{n}{\sigma_z^2} + \frac{m}{\sigma_y^2}(1+\eta) - \bar Y \frac{m}{ \sigma_y^2 + m \sigma_\theta^2} \\
  \bar Y \frac{m}{\sigma_y^2 } \\
  \eta \bar Y \frac{m}{\sigma_y^2 }
\end{bmatrix}.
\end{equation*}

The generative parameters described in the main text are as follows
\begin{Schunk}
\begin{Sinput}
> n=25 # Sample size for Z
> m=50 # Sample size for Y
> phi = 0
> theta = 1 # bias
> sigma_z = 2 # variance for Z
> sigma_y = 1 # variance for Y
\end{Sinput}
\end{Schunk}

The true bias is $\theta=1$. Assume we have an over-optimistic view of the bias, with prior distribution centered in 0 and relatively small prior variance.

\begin{figure}[!ht]
  \center
  \includegraphics[width=0.5\textwidth]{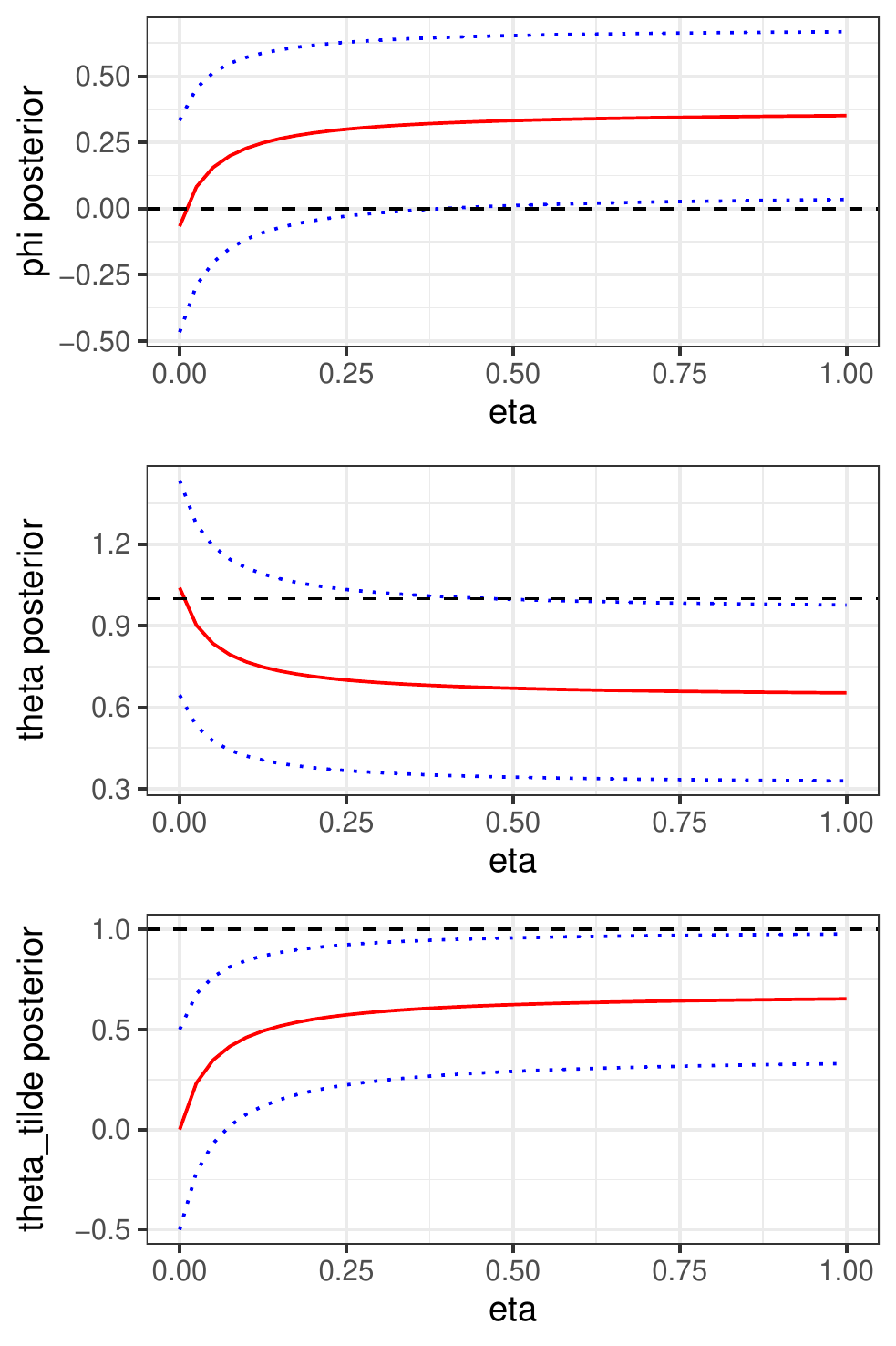}
  \caption{Posterior distribution of $\varphi$, $\theta$ and $\tilde\theta$ for a single dataset ($Z$,$Y$). A black horizontal line shows the true generative value. The posterior mean is the solid red line and we show intervals with $\pm$ one posterior std. dev. using dotted blue lines.}
  \label{fig:smi_post_5_1}
\end{figure}

In Figure~\ref{fig:smi_post_5_1} we show posterior distributions (mean $\pm$ std. dev.) for a randomly generated dataset ($\bar Z=$-0.0667; $\bar Y=$1.0562) using the generative parameters described in the main text. Note that the conventional bayes ($\eta=1$) is the worst estimation for the true parameters $\varphi$ accross all posible candidates $\eta \in [0,1]$.
\begin{Schunk}
\begin{Sinput}
> # Posterior for conventional bayes eta=1
> posterior = SMI_post_biased_data( Z=Z, Y=Y,
+                                   sigma_z=sigma_z, sigma_y=sigma_y,
+                                   sigma_phi=sigma_phi,
+                                   sigma_theta=sigma_theta, sigma_theta_tilde=sigma_theta,
+                                   eta=1 )
> posterior = mapply('rownames<-', posterior, MoreArgs=list(value=param_names))
> # posterior mean
> posterior$mean
\end{Sinput}
\begin{Soutput}
                 [,1]
phi         0.3511440
theta       0.6528197
theta_tilde 0.6528197
\end{Soutput}
\end{Schunk}

\subsubsection{Mean Square Error (MSE)}

From Equation~\ref{eq:smi_post}, we can compute the Posterior Squared Error (SE) of estimates arising from the SMI posterior $\varphi$, $\theta$ and , $\tilde\theta$,
\begin{align*}
  SE(\varphi) &= \Sigma_{[1,1]} + (\mu_{[1]}-\varphi^*)^2\\
  SE(\theta) &= \Sigma_{[2,2]} + (\mu_{[2]}-\theta^*)^2\\
  SE(\tilde\theta) &= \Sigma_{[3,3]} + (\mu_{[3]}-\theta^*)^2
\end{align*}

In the first simulation study of the main text we display the Mean Squared Error (MSE) for $\varphi$ and $\theta$, which is the result of averaging the posterior SE across datasets. We show that we can reach smaller MSE with values of $\eta$ other than 0 and 1. To generate these plots, we produced 1000 synthetic datasets, computed MSE using Eq.~\ref{eq:smi_post} on each one, with a grid of values of $\eta\in[0,1]$. The MSE lines displayed correspond to the \textit{average} MSE across datasets, for each value of $\eta$.

\begin{figure}[!ht]
  \center
  \includegraphics[width=0.5\textwidth]{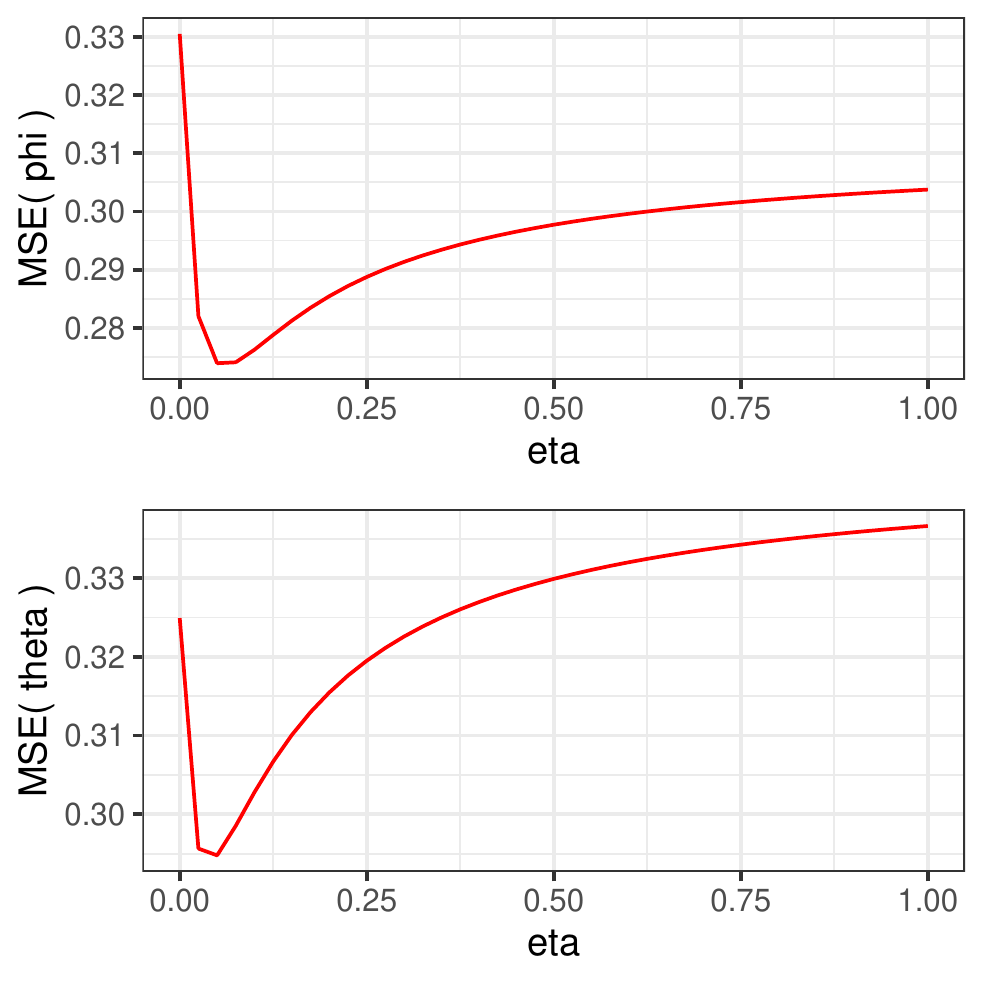}
  \caption{Mean Squared Error of the two main parameters under SMI posterior.}
  \label{fig:SMI_biased_MSE_average}
\end{figure}

\begin{figure}[!ht]
  \center
  \includegraphics[width=0.5\textwidth]{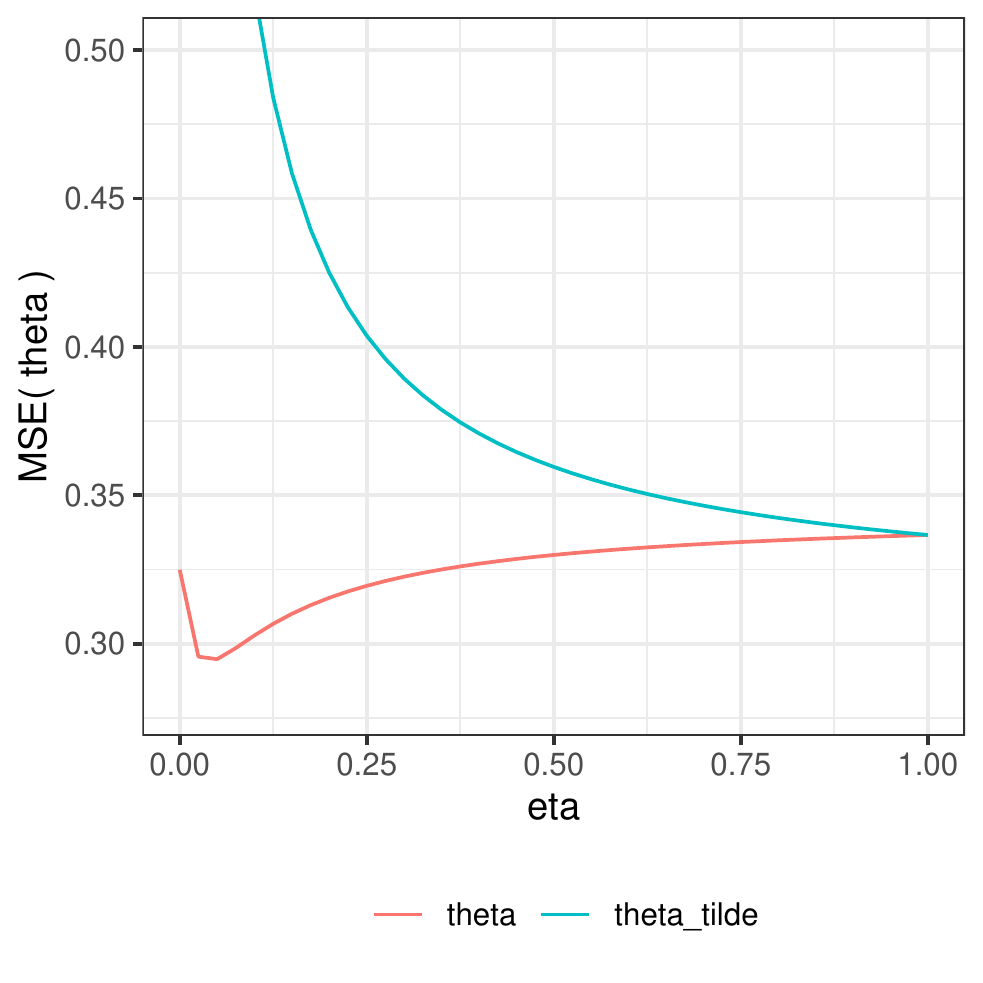}
  \caption{Comparison of the bias estimation under SMI (theta) vs powered likelihood (theta tilde)}
  \label{fig:SMI_biased_MSE_average_theta}
\end{figure}

Here we also show a comparison between $\theta$ and $\tilde\theta$. Figure~\ref{fig:SMI_biased_MSE_average_theta} shows that $\tilde\theta$ is dominated by $\theta$ in MSE. The comparison emphasises the convenience of SMI over power likelihood discussed in section 4.4 of the main text.

\subsubsection{Expected log pointwise predictive density (elpd)}

The elpd is 
\begin{equation*}
  elpd = \int\int p^*(z,y) \log p_{smi,\eta}( z, y \mid Z,Y) dz dy
\end{equation*}
where $p^*$ is the distribution representing the true data-generating process and
\begin{equation*}
  p_{smi,\eta}(z,y \mid Z,Y)=\int\int p(z,y \mid \varphi, \theta) \; p_{smi,\eta}(\varphi,\theta \mid Y,Z)\, d\varphi\,d\theta
\end{equation*}
is a candidate posterior predictive distribution, indexed by $\eta$.

Let $\begin{bmatrix}  a & b \\ b & c \end{bmatrix}=Cov(\varphi,\theta \mid Z,Y)^{-1}$ be the  inverse of the posterior covariance matrix of $(\varphi,\theta)$, and $\begin{bmatrix}  d \\ e \end{bmatrix}=E(\varphi,\theta \mid Z,Y)$ the posterior means.

Following straightforward Gaussian completion we can show that the joint posterior distribution for $\varphi$, $\theta$, and new data $z_0$ and $y_0$ is:
\begin{align*}
  p_{smi,\eta}(z_0,y_0,\varphi,\theta|Z,Y) &\propto p(z_0,y_0|\varphi,\theta) \; p_{smi,\eta}(\varphi,\theta|Z,Y) \\
  &\propto \exp\{ -\frac{1}{2} \; [ z_0^2 (\frac{1}{\sigma_z^2}) + y_0^2 (\frac{1}{\sigma_y^2}) + \varphi^2 (a+\frac{1}{\sigma_z^2}+\frac{1}{\sigma_y^2}) + \theta^2 (c + \frac{1}{\sigma_y^2} ) + \\
  & \hspace{2cm} - 2 z_0 \varphi (\frac{1}{\sigma_z^2}) - 2 y_0 \varphi (\frac{1}{\sigma_y^2}) - 2 y_0 \theta (\frac{1}{\sigma_y^2}) + 2 \varphi \theta (b+\frac{1}{\sigma_y^2}) \\
  & \hspace{2cm} -2 \varphi (a d + b e) -2 \theta (b d + c e ) ] \}
\end{align*}

So we have
\begin{equation} \label{eq:smi_post_pred_5_1}
p_{smi,\eta}(z_0,y_0,\varphi,\theta \mid Z, Y ) = \text{Normal}( \mu, \Sigma ),
\end{equation}

with
\begin{equation*}
  \Sigma = \begin{bmatrix}
  \frac{1}{\sigma_z^2} & 0 & -\frac{1}{\sigma_z^2} & 0 \\
  0 & \frac{1}{\sigma_y^2} & -\frac{1}{\sigma_y^2} & -\frac{1}{\sigma_y^2} \\
  -\frac{1}{\sigma_z^2} & -\frac{1}{\sigma_y^2} & a+\frac{1}{\sigma_z^2}+\frac{1}{\sigma_y^2} & b+\frac{1}{\sigma_y^2} \\
  0 & -\frac{1}{\sigma_y^2} & b+\frac{1}{\sigma_y^2} & c + \frac{1}{\sigma_y^2}
\end{bmatrix}^{-1} \text{, and } \mu = \Sigma \begin{bmatrix}
  0 \\
  0 \\
  a d + b e \\
  b d + c e
\end{bmatrix}.
\end{equation*}

We know the true generative values $\varphi^*$ and $\theta^*$, so we can compute $elpd$ using Monte Carlo samples from the true generative distribution $p^*$ and evaluate this values in the log-density of the bivariate normal $(z_0,y_0|Z,Y)$ from Equation~\ref{eq:smi_post_pred_5_1}.

\begin{figure}[!ht]
  \center
  \includegraphics[width=0.5\textwidth]{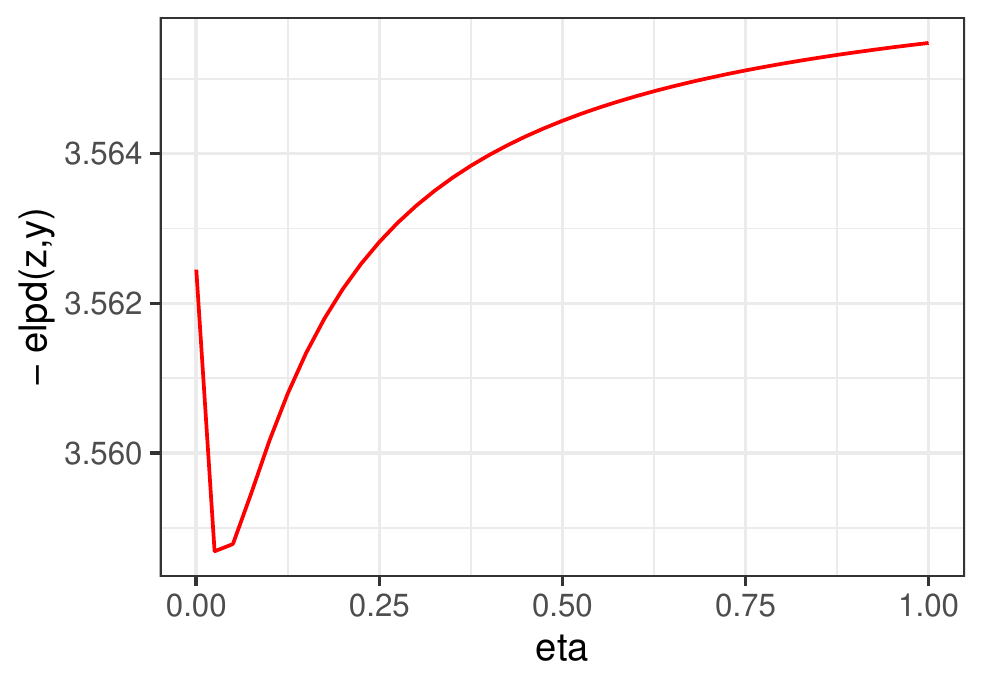}
  \caption{ELPD under SMI posterior.}
  \label{fig:SMI_biased_elpd}
\end{figure}

In Figure~\ref{fig:SMI_biased_elpd} we show the Monte Carlo estimation of the $elpd$. We select the optimal $\eta$ as the value that maximise the $elpd$. To generate this plot, we produced 1000 synthetic datasets, computed $elpd$ on each one (using Monte Carlo), with a grid of values of $\eta\in[0,1]$. The $elpd$ line correspond to the \textit{average} elpd across datasets, for each value of $\eta$.

\subsection{Agricultural data} \label{sec:agric_analysis}

The aim of the study, described in detail in \cite{Styring2017}, is to provide statistical evidence about a specific agricultural practice of the first urban centres in northern Mesopotamia.

The hypothesis is that increased agricultural production to support growing urban populations was achieved by cultivation of larger areas of land, entailing lower manure/midden inputs per unit area. This practice is known as \emph{extensification}.

Our contribution goes into extending the methods used to perform Bayesian analysis in this adverse scenario of model misspecification and big missing data.

\subsubsection{ Data }

The data consists of measurements of nitrogen and carbon isotopes for a collection of crop remains. There are two datasets: archaeological and modern, which we will denote by $\mathcal{A}$ and $\mathcal{M}$, respectively. First, the \emph{Archaeological} dataset $\mathcal{A}$, consists of data gathered from excavations of antique crop sites in the region of Mesopotamia. Second, the \emph{Modern} dataset, $\mathcal{M}$, was gathered in a controlled experimental setting in recent years. Further characteristics and description of variables on each dataset can be found in the statistical supplement to \cite{Styring2017}.

\subsubsection{ The model } \label{sec:agric_model}

\begin{figure}[!ht]
  \begin{center}
  \begin{tikzpicture}[thick,scale=1, every node/.style={transform shape}]
    \node (M_arc) [param] {$M_i$};

    \node (gamma) [param, fill=red!30, above=of M_arc, yshift=1.5cm, xshift=0.2cm] {$\gamma$};
    \node (alpha) [param, left=of gamma, xshift=0.2cm] {$\alpha$};
    \node (xi) [param, left=of alpha, xshift=0.2cm] {$\xi_s$};
    \node (sigma_xi) [param, left=of xi] {$\sigma_{\xi}$};
    \plate {po_rndeff} {(xi)} {\tiny $s \in S_{\mathcal{A}}$} ;

    \node (S_arc_po) [data, fill=blue!20, left=of M_arc, xshift=-0.5cm] {$S_i$};
    \node (P_arc_po) [data, above=of S_arc_po] {$P_i$};
    \plate {arc_po} {(P_arc_po) (S_arc_po) (M_arc)} {$i \in \mathcal{A}$};

    \node (C_arc_hm) [data, right=of M_arc] {$C_i$};
  	\node (P_arc_hm) [data, right=of C_arc_hm] {$P_i$};
  	\node (R_arc_hm) [param, below=of P_arc_hm] {$R_i$};
    \node (Z_arc_hm) [data, below=of C_arc_hm, yshift=-1cm] {$Z_i$};
    \plate {arc_hm} {(M_arc) (C_arc_hm) (P_arc_hm) (R_arc_hm) (Z_arc_hm)} {$i \in \mathcal{A}$};

    \node (beta) [param, right=of R_arc_hm, xshift=0.2cm] {$\beta$};
    \node (sigma) [param, below=of beta, yshift=0.2cm] {$\sigma$};
  	\node (v) [param, below=of sigma, yshift=0.2cm] {$v$};
    \node (zeta) [param, below=of v, yshift=0.2cm] {$\zeta_s$};
  	\node (sigma_zeta) [param, right=of zeta] {$\sigma_{\zeta}$};
    \plate {hm_rndeff} {(zeta)} {\tiny $s \in S_{\mathcal{A} \cup \mathcal{M}}$};

    \node (M_mod_hm) [data, right=of P_arc_hm, xshift=1.6cm] {$M_j$};
    \node (C_mod_hm) [data, right=of M_mod_hm ] {$C_j$};
  	\node (P_mod_hm) [data, right=of C_mod_hm] {$P_j$};
  	\node (R_mod_hm) [data, below=of P_mod_hm] {$R_j$};
    \node (Z_mod_hm) [data, below=of C_mod_hm, yshift=-1cm] {$Z_j$};
    \plate {mod_hm} {(M_mod_hm) (C_mod_hm) (P_mod_hm) (R_mod_hm) (Z_mod_hm)} {$j \in \mathcal{M}$};;

    \plate [dotted,label={ \large PO module}]{po} {(M_arc) (alpha) (gamma) (xi) (sigma_xi)  (S_arc_po) (P_arc_po) (po_rndeff) (arc_po) } {}
  	\plate [dotted, label={ \large HM module}]{hm} {(M_arc) (C_arc_hm) (P_arc_hm) (R_arc_hm) (Z_arc_hm) (arc_hm) (M_mod_hm) (C_mod_hm) (P_mod_hm) (R_mod_hm) (Z_mod_hm) (mod_hm) (beta) (sigma) (v) (zeta) (sigma_zeta) (hm_rndeff) } {}

  	\draw[dashed,red, line width=0.5mm] (-1.1,-1.2) to[out=90, in=180, distance=2cm] (1.1,1.3);

    \edge {sigma_xi} {xi};
    \edge {alpha, gamma, xi} {M_arc};
    \edge {S_arc_po, P_arc_po} {M_arc};

    \edge {sigma_zeta} {zeta};
    \edge {beta,zeta,sigma,v} {Z_arc_hm};
    \edge {beta,zeta,sigma,v} {Z_mod_hm};
  	\edge {M_arc, C_arc_hm, P_arc_hm, R_arc_hm} {Z_arc_hm};
    \edge {M_mod_hm, C_mod_hm, P_mod_hm, R_mod_hm} {Z_mod_hm};

  \end{tikzpicture}
  \end{center}
  \caption[Graphical model for agricultural data]{ Graphical representation of the model for the agricultural data. Squares denote observable variables and circles denote unknown quantities (parameters and missing data).The main interest of the study is on the parameter $\gamma$ (red circle), effect of size $S_i$ (blue square) on Manure level $M_i$. The dashed line indicates the cut where SMI is applied for the imputation of missing manure.}
  \label{fig:agricurb_model}
\end{figure}
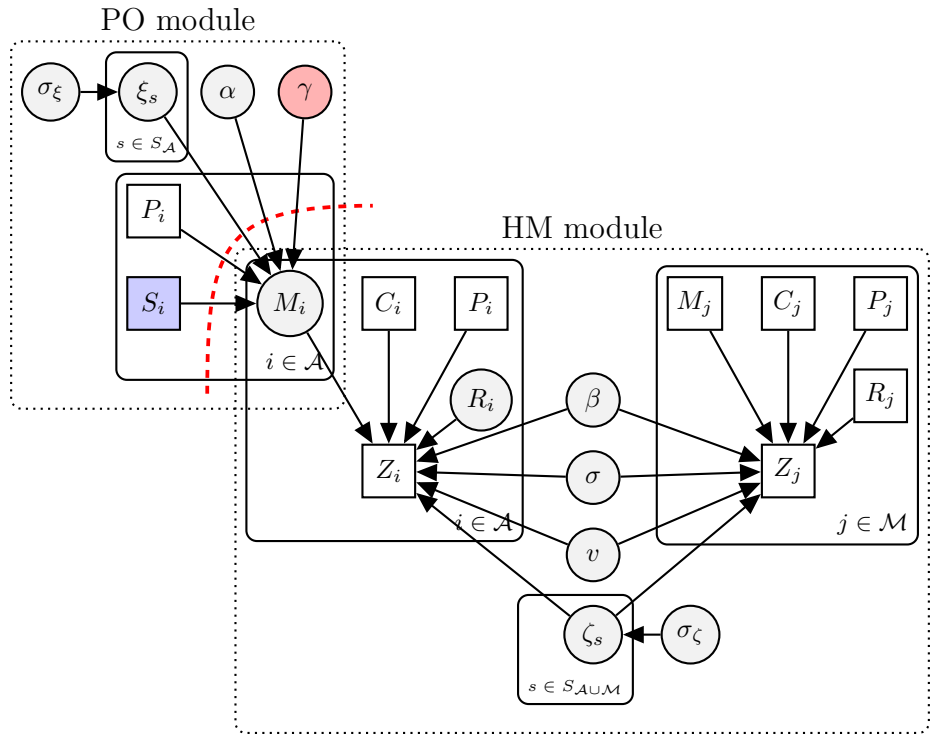

We preserved the model stated in \cite{Styring2017}. See Figure~\ref{fig:agricurb_model}. The main goal of that study is to test the hypothesis of \emph{extensification} in the ancient urban sites. This hypothesis can be condensed in analysing the strength of the effect of the site size $S_i$ on the corresponding manure level $M_i$ for the records in the archaeological data, $\{i \in \mathcal{A}\}$. The more negative the estimated effect, the stronger the evidence supporting the extensification hypothesis.

However, there is no available information about the Manuring levels in the ancient dataset. This is addressed by using a \emph{Data augmentation} perspective, and consider that all the corresponding values of the manure level in the Archaeological data are missing.

The model consists of a two-module model, which integrates archaeological and calibration data so the missing manuring levels can be inferred. The first module is a Proportional Odds model ($PO$), with the missing Manure Levels in the archaeological data as the ordinal response ($M_i; i \in \mathcal{A}$). The second module consists of a linear Gaussian model ($HM$), applicable to both datasets, with the Nitrogen level of the crops ($Z$) as the response, and Manure levels as one of the predcitors. The graphical representation of the model is depicted in Figure~\ref{fig:agricurb_model}

The model can be studied from a multi-modular perspective. Indeed, the supplement of \cite{Styring2017} performs Bayesian Multiple Imputation (BMI) to impute the missing manure levels with parameters learnt from the $HM$ module, and therefore \emph{cutting} the influence from the $PO$ module into this imputation. The parameters of the $PO$ module are then inferred conditional on the imputed values and other information in the archaeological data.

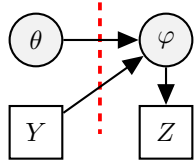
\begin{figure}[!ht]
  \begin{center}
  \begin{tikzpicture}[thick,scale=1, every node/.style={transform shape}]
    \node (theta) [param] {$\theta$};
    \node (varphi) [param, right=of theta, xshift=0.5cm] {$\varphi$};
    \node (Y) [data, below=of theta] {$Y$};
    \node (Z) [data, below=of varphi] {$Z$};

  	\draw[dashed, red, line width=0.5mm] (0.85,0.5) to (0.85,-1.25);

    \edge {theta, Y} {varphi};
    \edge {varphi} {Z};

  \end{tikzpicture}
  \end{center}
  \caption{ Simplified representation of the model for the agricultural data.}
  \label{fig:agricurb_model_simple}
\end{figure}

In Figure~\ref{fig:agricurb_model_simple} we draw a simplified version of the complete model to clarify how the SMI framework can be applied in this setting. The mapping of variables between the complete and simplified graphs is as follows: the missing manure levels will take the role of $\phi$; the observed data in the $HM$ module is $Z$, the parameters in the $PO$ module will be $\theta$, and the rest of archaeological data in the $PO$ module is $Y$.

The simplified version resembles our graphical model in Figure~\ref{fig:toy_multimodular_model}. From here it is clear how the Bayesian imputation approach taken by \cite{Styring2017} is equivalent to a cut model: first learn $\phi$/manure-level from the $HM$ module, and then learning $\theta$ (ie $\gamma$) conditional on $\phi$ and $Y$. This scheme yields our known cut posterior from Eq.~\ref{eq:cut_post}.
\[P_{cut}(\varphi,\theta \mid Z,Y) = P(\varphi \mid Z) \; P(\theta \mid Y,\varphi) \]

We extend the Bayesian imputation approach in \cite{Styring2017} and apply Semi-Modular Inference to their setting.
\[P_{smi,\eta}(\varphi,\theta \mid Z,Y) = \int P_{pow,\eta}(\varphi, \tilde\theta \mid Z, Y) \; P(\theta \mid Y,\varphi) d\tilde\theta \]

SMI allows us to expand the space of candidate posteriors in a way that we can control smoothly, rather that eliminating the influence of the $PO$ module in the imputation of the missing Manure Levels.

\begin{figure}[!ht]
  \center
  \includegraphics[width=0.7\textwidth]{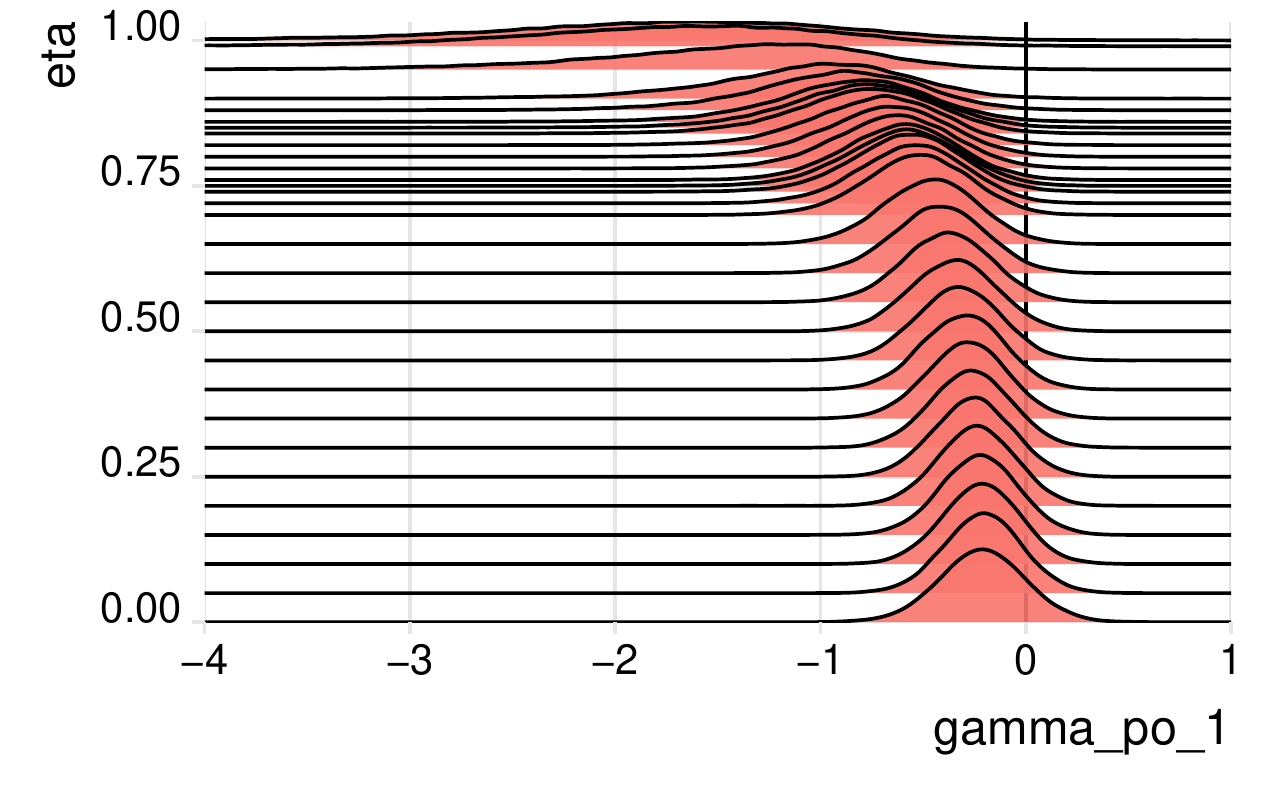}
  \caption{SMI posterior distribution for the parameter of interest in the agricultural model.}
  \label{fig:agric_smi_post_gamma}
\end{figure}

In Figure~\ref{fig:agric_smi_post_gamma} we display the collection of candidate distributions spanned by SMI by considering a grid of values for $\eta\in[0,1]$. The distribution at $\eta=$0 is comparable to the Bayesian Imputation approach in \cite{Styring2017}.

\begin{figure}[!ht]
  \center
  \includegraphics[width=0.7\textwidth]{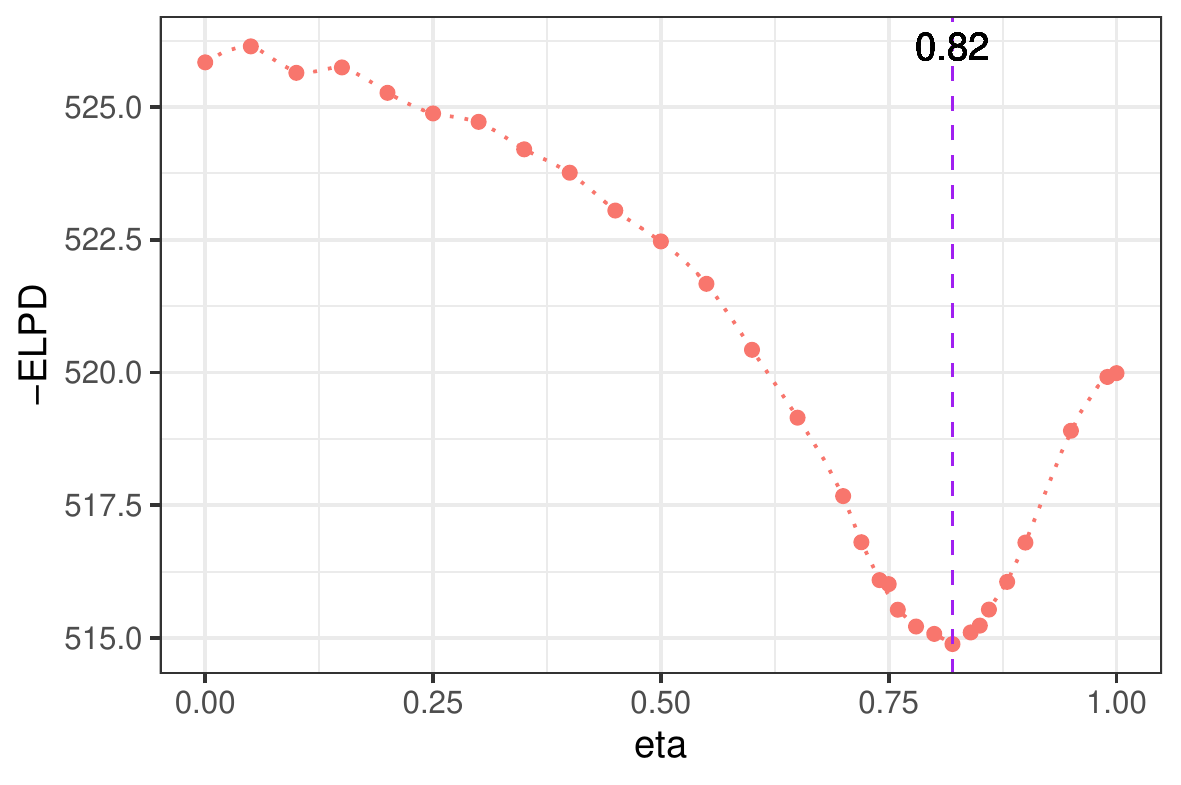}
  \caption{Choosing the best SMI posterior candidate by choosing the $\eta$-value maximizing the ELPD.}
  \label{fig:agric_smi_model_eval_elpd_waic}
\end{figure}

To choose the optimal value, $\eta^*$ say, we maximise the Expected Log Predictive Density or \emph{ELPD} \citep{Vehtari2016} to predict a new value of the response in the HM module ($Z_i; i \in \mathcal{A}$ in the diagram). In Figure~\ref{fig:agric_smi_model_eval_elpd_waic} we show the negative ELPD as a function of $\eta$. The optimal value is reached at $\eta^*=$0.82.

\begin{figure}[!ht]
  \center
  \includegraphics[width=0.7\textwidth]{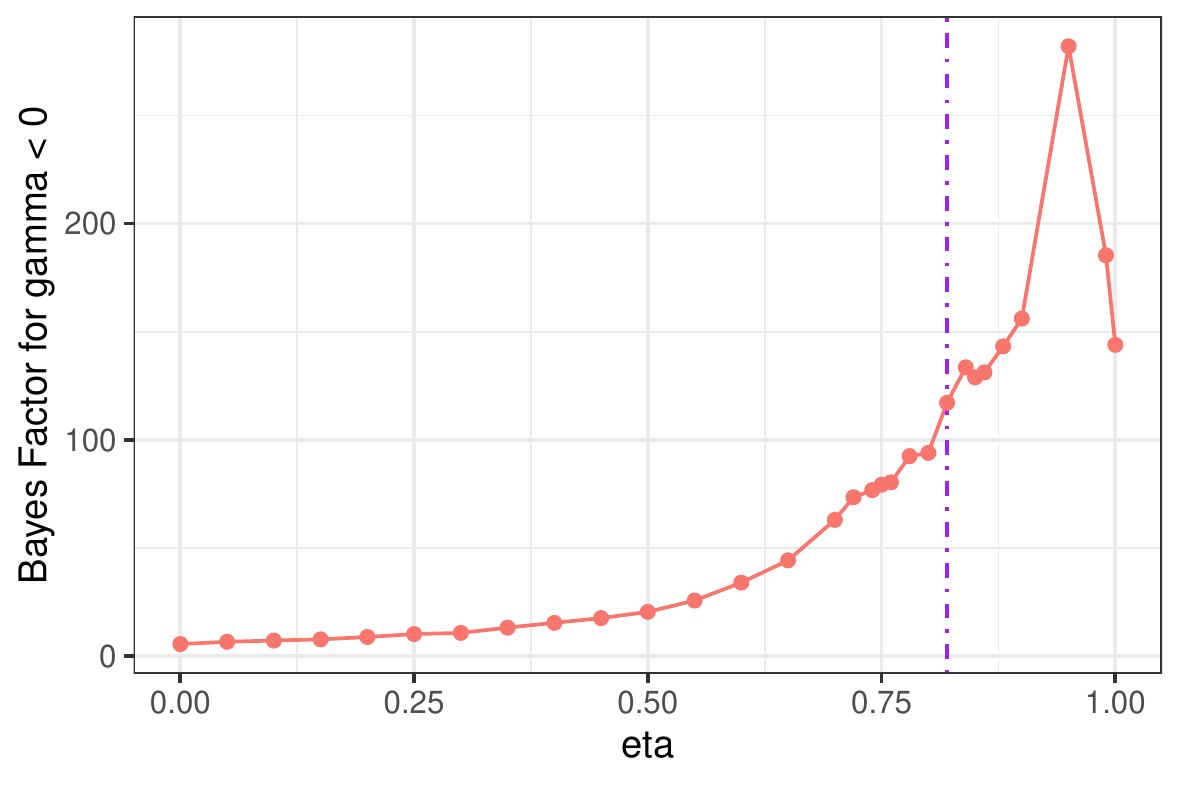}
  \caption{Bayes Factor for the hypothesis $\gamma \leq 0$ across all values of $\eta$.}
  \label{fig:agric_smi_gamma_leq_0_BF}
\end{figure}

The elpd changes dramatically over the range $0\le \eta\le 1$ and $\eta^*$ is clearly distinguished from 0 or 1. This impacts downstream inference: the Bayes factor of interest changes from $BF_{smi,0}(\gamma \leq 0)=$5.54 in the cut model, to $BF_{smi,0.82}(\gamma \leq 0)=$117.11 in the optimal SMI posterior with $\eta=$0.82, a substantial shift in the strength of the evidence for the \emph{extensification} hypothesis. In Figure~\ref{fig:agric_smi_gamma_leq_0_BF} we show the comparison of Bayes Factors  accross values of $\eta \in [0,1]$ for the hypothesis $\gamma \leq 0$.

\clearpage
\newpage

\bibliographystyle{apalike} 
\bibliography{references} 